\documentclass[acmsmall]{acmart}

\usepackage{tikz}
\usetikzlibrary{arrows,shapes,positioning,shadows,trees}
\tikzset{
  basic/.style  = {draw, text width=8cm,  font=\sffamily, rectangle, align=center},
  root/.style   = {basic, rounded corners=2pt, thin, align=center,   },
  level 2/.style = {basic,  thin,align=center,  text width=10em},
  level 3/.style = {basic, thin, align=left, text width=10em}
}
\usepackage[edges]{forest}
\usetikzlibrary{arrows.meta}
\usepackage{bbding}
\usepackage{float}
\usepackage{amsmath}
\usepackage{amstext}
\usepackage{geometry}
\usepackage{xr}
\usepackage{svg}
\usepackage{pifont}
\usepackage{wasysym}
\usepackage{lscape}
\usepackage[super]{nth}
\usepackage{subfig}
\usepackage{fancyvrb,cprotect}
\usepackage{enumerate}
\usepackage{array, longtable}
\newcommand{\xdownarrow}[1]{%
  {\left\downarrow\vbox to #1{}\right.\kern-\nulldelimiterspace}
}
\usepackage[T1]{fontenc}
\usepackage[utf8]{inputenc}
\usepackage[font=small,labelfont=bf]{caption}
\usepackage{booktabs}
\usepackage{varwidth}
\usepackage{multirow}
\newsavebox\tmpbox
\AtBeginDocument{%
  \providecommand\BibTeX{{%
    \normalfont B\kern-0.5em{\scshape i\kern-0.25em b}\kern-0.8em\TeX}}}

\setcopyright{acmcopyright}
\copyrightyear{2021}
\acmYear{2021}
\acmDOI{10.1145/1122445.1122456}

\acmJournal{CSUR}
\acmVolume{37}
\acmNumber{4}
\acmArticle{111}
\acmMonth{1}

\makeatletter
\newcommand*{\addFileDependency}[1]{
  \typeout{(#1)}
  \@addtofilelist{#1}
  \IfFileExists{#1}{}{\typeout{No file #1.}}
}
\makeatother

\newcommand*{\myexternaldocument}[1]{%
    \externaldocument{#1}%
    \addFileDependency{#1.tex}%
    \addFileDependency{#1.aux}%
}
\myexternaldocument{aaaaa}
\begin{document}

\title{Social Media Identity Deception Detection: A Survey }

\author{Ahmed Alharbi}

\email{ahmed.alharbi@rmit.edu.au}
\author{Hai Dong}
\email{hai.dong@rmit.edu.au}
\author{Xun Yi}
\email{xun.yi@rmit.edu.au}
\author{Zahir Tari}
\email{zahir.tari@rmit.edu.au}
\author{Ibrahim Khalil}
\email{ibrahim.khalil@rmit.edu.au}
\affiliation{%
  \institution{School of Computing Technologies, RMIT University}
  \city{Melbourne}
  \country{Australia}
}

\renewcommand{\shortauthors}{ A. Alharbi et al.}

\begin{abstract}
Social media have been growing rapidly and become essential elements of many people's lives. Meanwhile, social media have also come to be a popular source for identity deception. Many social media identity deception cases have arisen over the past few years. Recent studies have been conducted to prevent and detect identity deception. This survey analyses various identity deception attacks, which can be categorized into fake profile, identity theft and identity cloning. This survey provides a detailed review of social media identity deception detection techniques. It also identifies primary research challenges and issues in the existing detection techniques. This article is expected to benefit both researchers and social media providers.
\end{abstract}



\maketitle

\section{Introduction}
Social media are web-based communication tools that facilitate people with creating customized user profiles to interact and share information with each other. Social media have been growing rapidly in the last two decades and have become an integral part of many people's lives. Users in various social media platforms totalled almost 3.8 billion at the start of 2020, with a total growth of 321 million worldwide in the past 12 months\footnotemark. Users spend roughly 15\% of their waking lives on social media. Facebook surpasses other social media platforms in terms of popularity. There are approximately 2 billion users on Facebook, followed by YouTube with 1.9 billion users.
\footnotetext{https://wearesocial.com/blog/2020/01/digital-2020-3-8-billion-people-use-social-media}

Trust has been well acknowledged as a decisive aspect for the success of social media platforms \cite{leung2013social}\cite{Zhang2018a}. In a survey conducted in 2017 that involved 9,000 users, 40\% of respondents reported that they deleted their social media accounts because they did not trust that the platforms can protect their private information. 53\% of the online users concerned about online privacy in contrast to a year ago as of February  2019\footnote{https://www.statista.com/statistics/373322/global-opinion-concern-online-privacy/}. 81\% of the online users in the United States felt that their personal information is vulnerable to hackers as of July 2019\footnote{https://www.statista.com/statistics/972911/adults-feel-data-personal-information-vulnerable-hackers-usa/}. Almost 16.7 million United States citizens were victims of identity theft in 2017\footnote{https://www.javelinstrategy.com/press-release/identity-fraud-hits-all-time-high-167-million-us-victims-2017-according-new-javelin}. 

Social media platforms can be classified according to the presence and richness of content, self-representation and self-disclosure \cite{kaplan2010users}\cite{tsikerdekis2014online}. The presence of content is influenced by the immediacy and intimacy of social media communication, while the richness of content refers to the total amount of information which can be carried on social media \cite{daft1986organizational}\cite{short1976social}. Self-representation indicates the self-control of users' expression, while self-disclosure reveals whether users' published information is voluntarily or not \cite{goffman1978presentation}\cite{kaplan2010users}. The above characteristics classify social media platforms into eight categories, as shown in Table \ref{soaC}. They consist of collaborative projects, blogs, microblogging, virtual game worlds, content communities, social networking sites, social news sites,  and virtual social worlds \cite{kaplan2010users}\cite{tsikerdekis2014online}. This article focuses on these general types of social media platforms.
\begin{table}[]
\caption{Social media platforms classifications}
\label{soaC}
\resizebox{0.9\textwidth}{!}{%
\begin{tabular}{|m{0.4\textwidth}|c|m{1.8cm}|m{1.9cm}|m{2cm}|m{1.8cm}|}
\hline
\multirow{2}{*}{}                                      & \multicolumn{5}{c|}{\textbf{Presence and richness of content}}                                                                                                               \\ \cline{2-6} 
                                                       & \multicolumn{5}{l|}{Low $\xrightarrow{\hspace*{8cm}}$ High}                                                                                                                                       \\ \hline
\multirow{2}{*}{\textbf{Self-representation / Self-disclosure}} & \multirow{2}{*}{\begin{tabular}[c]{@{}l@{}}Low \\ $\xdownarrow{0.5cm}$ \\ High\end{tabular}} & Collaborative  projects & Social news sites & Content communities     & Virtual game worlds   \\ \cline{3-6} 
                                                       &                                                                     & Blogs                   & Microblogging     & Social networking sites & Virtual social worlds \\ \hline
\end{tabular}
}
\end{table}

People and organizations have an incentive to take advantage of social media platforms. A certain type of propaganda can be published to obtain an unrealistic level of influence \cite{silawan2017sybilvote}. It is easy to specify a certain group on social media platforms and attack them based on a prepared strategy. In these days, it has been a popular phenomenon that social media users purchase Twitter followers, Facebook likes, Amazon reviews and YouTube comments. These services can be achieved through strategies such as creating multiple fake profiles, employing compromised accounts, or even paying users to post content over their accounts \cite{chen2016statistical}\cite{yang2016leveraging}.

\begin{figure*}%
    \centering
    \includegraphics[scale=0.15]{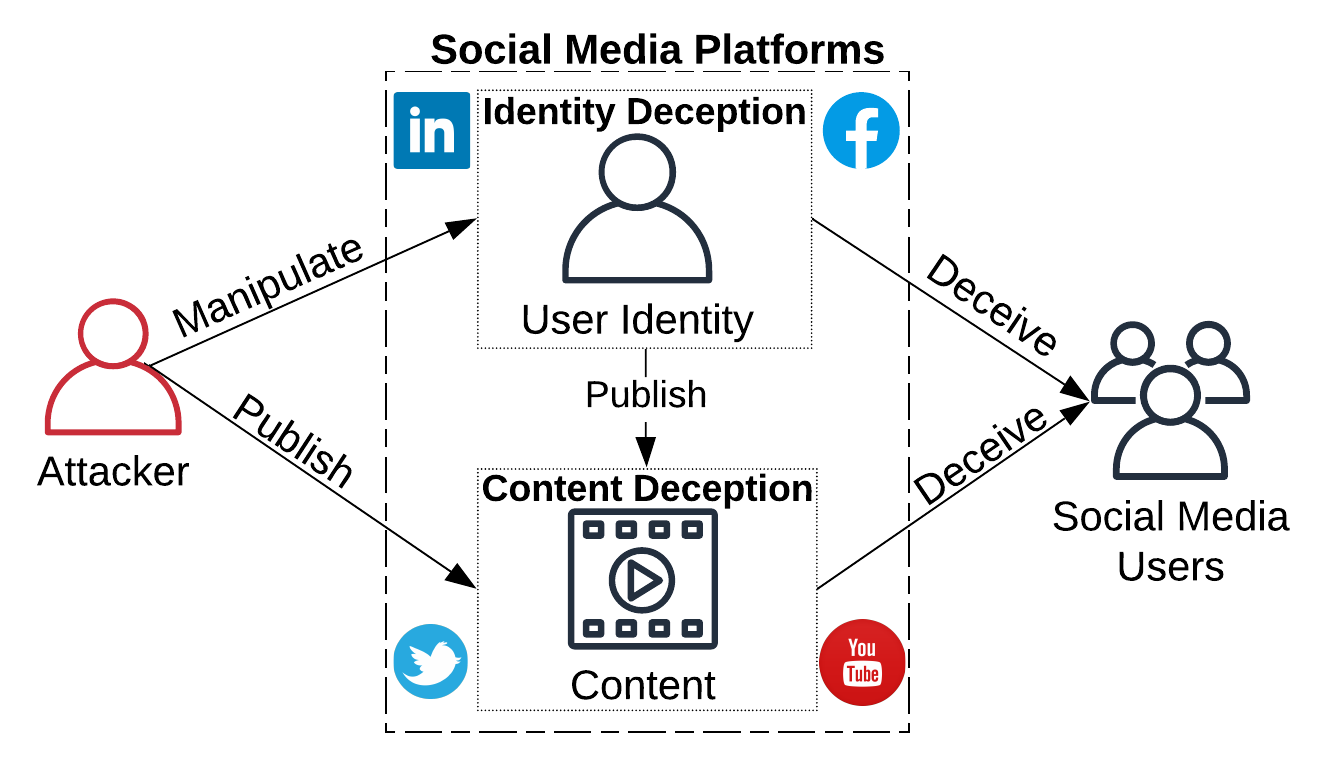} 
    \caption{Social media deception}%
    \label{DEC}%
\end{figure*}

Deception in social media can be divided into content deception and identity deception \cite{madhusudan2003text}\cite{Tsikerdekis2015}, as shown in Figure \ref{DEC}. The content deception is to manipulate social media content by tampering with images, spreading spams and sending malicious links. The content deception mostly happens on social media platforms, such as social news sites and blogs. The identity deception is to manipulate the user's identity information or to impersonate someone's identity to deceive social media users. Successful identity deception is achieved when the user's personal information is not verified \cite{Tsikerdekis2015}.

Social media identity deception can easily be achieved. Social media only require a valid email address to register a new account and create a personalized profile (identity) \cite{Bahri2018}. The main issue is that the information given in the registration procedure does not subject to any verification process. Therefore, a user can pretend to be someone else with the intention of deceiving other users. For example, a malicious account can pretend to be Alice who is Bob's close friend. Alice sends a friendship invitation to Bob. As a result, the malicious account will obtain access to Bob's private information that Bob intends to share only with the real Alice. Furthermore, fake identities (profiles) tend to commit other malicious behaviours. For instance, fake identities can attack the vulnerabilities of social media, spread malware and manipulate legitimate users.

Several existing surveys have been conducted in the relevant fields.
Kaur et al. \cite{Kaur2018} reviewed spammers and compromised account detection techniques. Masood et al. \cite{Masood2019}  conducted a survey on various approaches of spam detection on Twitter. Velayudhan and Somasundaram \cite{pcompromised} examined various methods for compromised account detection. Further, Al-Qurishi et al. \cite{Al-qurishi2017} reviewed Sybil attacks detection techniques. Ramalingam et al. \cite{Ramalingam2018} also conducted a review of Sybil detection approaches. Finally, Bahri et al. \cite{Bahri2018}  provided a review of identity management processes, identity attacks and defence mechanism. These surveys have focused especially on single categories of identity deception, such as Sybil attack and compromised accounts. None of them have focused on comprehensively analysing identity deception attacks on social media.

This article is motivated by the massive growth of social media usage. Social media have become a popular tool for identity deception. Social media identity deception cases frequently occurred in recent years. Table \ref{Comp_real_identity_attacks} lists several typical cases. There is an urgent need to detect and prevent those threats. Furthermore, social media identity deception detection is still in its infancy, with several open research issues and challenges remaining to be investigated. It is important to undertake further research that can improve social media identity deception techniques. This article analyses the identity deception attacks on social media platforms. This article then reviews identity deception detection techniques in detail.  This article makes the following contributions:
\begin{enumerate}[i]
    \item It presents an overview of social media identity deception attacks available in the literature, followed by a comparison between each identity deception attack in social media platforms. A high-level analysis of identity deception attacks on social media platforms has been conducted covering fake profile, identity theft and identity cloning. Each identity deception has been critically examined along with a use case scenario.
    \item It provides a comprehensive analysis of existing identity deception detection techniques in social media platforms. An exhaustive survey has been conducted focusing on up-to-date detection techniques. Each identity deception detection technique has been critically analysed and compared in five dimensions: domain, detection techniques, features, evaluation metrics and datasets.
    \item It summarizes various open researches and provides future directions for identity deception detection techniques in social media based on this literature review. Open research challenges are grouped into general and specific research challenges.  The general research challenges present the open research challenges shared among all the identity deception detection techniques, while the specific research challenges present the open research challenges to the particular subcategories of identity deception detection techniques. 
\end{enumerate}

The survey is organized as follows. Section 2 introduces the previous surveys on social media identity deception detection. Section 3 introduces the social media identity deception attacks. Section 4 reviews the detection techniques of social media identity deception. Section 5 discusses and compares identity deception detection techniques in five dimensions. Section 6 summarizes the identified research gaps and the future research directions.  Section 7 concludes this article.

\begin{table*}[]
\centering
\caption{Real-life cases of social media threats}
\label{Comp_real_identity_attacks}
\resizebox{\textwidth}{!}{%
\smallskip 
\begin{tabular}{lp{0.15\textwidth}p{0.15\textwidth}p{0.40\textwidth}p{0.35\textwidth}}
\toprule
\textbf{Year}                            & \textbf{Social Media} & \textbf{Tactic}& \textbf{Approach} & \textbf{Impact}   \\ \toprule
July 2017 \cite{counter}  & LinkedIn, Facebook, Blogger and WhatsApp & Fake Profile & Adversaries created a fake identity who was a U.S.-based photographer named Mia Ash. The adversaries spread a Remote Access Trojan (RAT), called PupyRAT, via these social media honeypot accounts to hijack the controls of victims’ devices. & The fake account hacked several accounts across social networks.\\ [-0.70em] \\

March 2017 \cite{russell_2017} & Twitter & Compromised account & A Turkish attacker took advantage of a 3rd-party app vulnerability called TwitterCounter to hack Amnesty International, Forbes and other prominent organizations. \newline Attacker tweeted fake news to the Netherlands. & Posted tweets were deleted.\\ [-0.70em]\\

Early 2017 \cite{calabresi_2017}  & Twitter  & Phishing/ Malware and Fraudulent Accounts & 10000 custom phishing messages were sent by Russian operatives to Twitter users in the Defence Department. \newline Every link was bugged with malware which enables the adversary to control and access the victim's system.
  & This attack represented a major development in cyber capabilities. \newline  This attack was considered to be an escalation in Russia’s cyberwar against the US and was considered to be the most coordinated attack and was well-organized.  \\ [-0.70em]\\
  
July 2015 \cite{intelligence_2015} &	Twitter & Malware/ Data Exfiltration &   HAMMERTOSS ruined network defenders’ strength by mimicking the behaviours of benign accounts. It uses several commonly visited websites such as GitHub and Twitter to relay commands and extract data from victims. 
& Developers of HAMMERTOSS  performed an attack against the Joint Chiefs of Staff and the White House. \\ [-0.70em]\\
  
April 2013 \cite{fisher_2019} & Twitter &	Compromised account & Associated Press account was compromised, and posted fake breaking news that the White House was exploding.  & The Dow thereafter declined about 150 points before stabilizing. \$136 billion in economic value were erased because of the fake news. \\ 
\toprule
\end{tabular}%
}

\end{table*}
\section{Previous Surveys on Social Media Identity Deception}
It is unsurprising that there are already surveys conducted on social media identity deception detection. This section summarizes those surveys and states the major difference of our survey.

Al-Qurishi et al. \cite{Al-qurishi2017} reviewed detection and prevention techniques for Sybil attacks in social networks. Sybil attacks detection techniques were categorized into graph, machine learning,  manual verification and prevention-based classes. Graph-based techniques use the information on social networks to minimize Sybil attacks, 
which is mostly based on Random Walk (RW). Machine-learning-based techniques are designed to solve problems involving an enormous quantity of data. Manual verification bases on social media content. Prevention approaches are intended to prevent the creation of Sybil accounts.

Bahri et al. \cite{Bahri2018} performed a review on identity management processes, identity attacks and defence mechanisms. Social media identity attacks were divided into target attacks (to specific users) and mass attacks (to whole social media platforms). The target attacks were classified into identity theft and cloning attacks. 
This article also discusses mass attacks defence mechanisms.

Kaur et al. \cite{Kaur2018} conducted a survey on detection of spammers/spam accounts and compromised accounts. The authors split spam detection into the following types: honey profiles, URL/blacklist, machine learning classification techniques, incremental learning and clustering techniques. The authors then grouped the dimensions of compromised account detection into content, message characteristics, graph structure, temporal characteristics and other complementary analysis. 

Masood et al. \cite{Masood2019} conducted a survey on various techniques of spammer detection on Twitter. The authors classified the detection approaches into fake content based, URL based, intending topics based and fake user identification based.

Velayudhan and Somasundaram \cite{pcompromised} conducted a survey on the various methods for compromised account detection in social networks. The authors categorized these detection techniques into seven main classes: text mining, statistical techniques, role-driven behavioural diversity, time profile, behavioural profile, cross-platform and composite behavioural modelling.

Ramalingam et al. \cite{Ramalingam2018} reviewed Sybil detection approaches in social networks. These approaches were grouped into content-based defence, graph-based defence and hybrid approaches. The works of Sybil detection were mostly realized by using the social graph-based defence. This type of defence considers nodes (users) features and links (edges/relations) between nodes.

Table \ref{COMP} shows the major differences between the scopes of our survey and the existing surveys. None of the existing surveys have comprehensively investigated all the available types of identity deception attacks and their corresponding detection technologies. Most of the existing surveys only focus on single types of identity deception attacks. There lacks an in-depth insight into the broad area of identity deception attacks and the existing detection technologies as well as the challenges they encounter. The purpose of this article is to present an overview of most identity deception attacks on social media platforms covering an in-depth analysis of each identity deception attack and a comprehensive analysis of each type of identity deception attack detection techniques.


\begin{table*}[]
\centering
\caption{Comparison between our survey and the existing surveys}
\label{COMP}
\resizebox{0.8\textwidth}{!}{%
\smallskip 
\begin{tabular}{p{2.5cm}p{2.5cm}p{0.05\textwidth}p{0.05\textwidth}p{0.05\textwidth}p{0.05\textwidth}p{0.05\textwidth}p{0.05\textwidth}p{2cm}}
\toprule
\textbf{Attack}                            & \textbf{Sub-attack} & \textbf{\cite{Kaur2018}} & \textbf{\cite{Bahri2018}}  & \textbf{\cite{pcompromised}} &  \textbf{\cite{Al-qurishi2017}} & \textbf{\cite{Masood2019}}& \textbf{\cite{Ramalingam2018}} &    \textbf{Our survey}   \\ \toprule
Fake Profile     & Sybil attack  & & \textbf{\checkmark} &  & \textbf{\checkmark} && \textbf{\checkmark} &   \textbf{\checkmark} \\ 
                                           & Sockpuppet    &  &   &&&&&\textbf{\checkmark}       \\ 
                                           & Social botnet   & & &&&&&\textbf{\checkmark}    \\ 
                                           & General   & & &&&\textbf{\checkmark}&& \textbf{\checkmark}    \\
\toprule
Identity Theft   & Compromised  & \textbf{\checkmark} & & \textbf{\checkmark} &&&& \textbf{\checkmark}\\   
\toprule
Identity Cloning & &  &   &&&&& \textbf{\checkmark}      \\ \toprule
                                           
\end{tabular}%
}
\end{table*}

\section{Social Media Identity Deception Attacks}
Social media are designed to be interaction-based applications. Social media allow users to share the way they live with their friends and others over the Internet. The users of social media produce enormous data. The users' sensitive information posted on social media increases the risk of cyberattacks \cite{Fire2014a}. A user only needs to provide a valid email to register with most social media platforms. It is easy to register with a forged identity which does not need to correspond to the identity of a real person. The information provided in the registration process usually does not subject to rigorous validation mechanisms. This enables deceivers to create multiple social media accounts, and to impersonate anyone they target. 

\begin{figure}
\centering
\begin{forest}
  for tree={
    align=center,
    font=\scriptsize,
    edge+={thin, -{Stealth[]}},
  },
  forked edges,
  if level=0{
    inner xsep=1pt,
    tikz={(.children first) -- (.children last);}
  }{},
  [\textbf{Social Media Identity Deception Attacks}
    [Fake Profile (Sec. \ref{fakeprofile})
    [Sybil \\Attack (Sec. \ref{sybilattack})]
    [Sockpuppets \\ (Sec. \ref{sockpp})]
    [Social \\Botnet (Sec. \ref{socialbot})]
    ]
    [Identity Theft (Sec. \ref{identitytheft})
    [Compromised \\Account (Sec. \ref{compacc})]
    [Hijacked \\Account (Sec. \ref{hijacc})]]
    [Identity Cloning (Sec. \ref{identityclonin})
    [Single-site \\ (Sec. \ref{sinsite})]
    [Cross-site  \\ (Sec. \ref{crosssite})]]
  ]
\end{forest}
\caption{Identity deception attacks on social media platforms} \label{identity_attacks}
\end{figure}
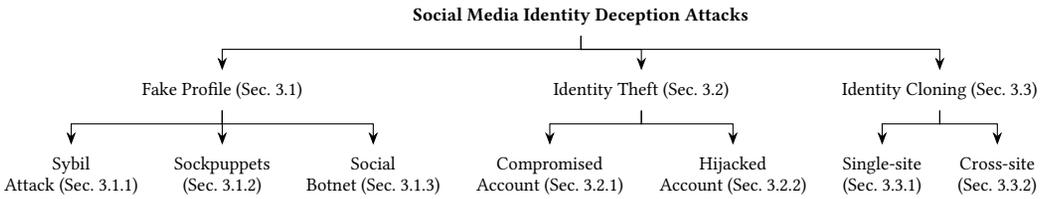

Figure \ref{identity_attacks} shows a taxonomy of identity deception attacks on social media platforms.
We classify the social media identity deception attacks into fake profile, identity theft and identity cloning. We discuss those categories and subcategories of social media identity deception attacks in the rest of the section. The other fake profile is a mixture of many minor fake profile categories. We do not provide the details of the other category in this section. Instead, we explain them in Section 4.1.4.

\subsection{Fake Profile}\label{fakeprofile}
Fake profiles can be classified into malicious and non-malicious accounts. Non-malicious accounts create a fake profile simply with the purpose of having multiple accounts \cite{Kaur2018}. In contrast, malicious accounts spread spam content among real users and perform phishing attacks. Facebook removed a whopping 2 billion fake accounts in the \nth{1} quarter of 2019\footnote{https://transparency.facebook.com/community-standards-enforcement}. It has been announced that 70 million fake accounts on Twitter were suspended in May and June 2017\footnote{https://www.engadget.com/2018/07/06/twitter-reportedly-suspended-70-million-fake-accounts-in-may-and/}. There are various types of fake profile on social media platforms. We describe the most common types in the following subsections.

\subsubsection{Sybil Attacks}\label{sybilattack}
\paragraph{Definition:}
In computer security, Sybil attacks occur when an attacker creates a large number of pseudonymous profiles (identities), which are called Sybil identities, as shown in Figure \ref{Fake_profile}a. These profiles intend to significantly impact or subvert a system \cite{borgatti2006graph}\cite{ douceur2002sybil}. In social media platforms, a Sybil identity is a fake profile with which an attacker attempts to register a large number of fake identities to befriend with many benign users. A Sybil identity can conduct various suspicious activities in social media platforms. An example of such an activity is to promote the reputation and popularity of a user in e-commerce settings by voting the target user as ``good''.
\paragraph{Use Case:}
An attacker registers a large number of pseudonymous profiles, aiming to attain significant influence in a social network. The adversary can strategically initiate befriending and defriending 
actions by taking advantages of the multiple fake profiles. Therefore, the adversary can manipulate the social graph neighbourhood topology which is controlled by the number of nodes mastered by the adversary. The adversary can deceive other users of the social network or even fool the social network administration. 
\subsubsection{Sockpuppet}\label{sockpp}
\begin{figure}%
    \centering
    \subfloat[Sybil attack]{{\includegraphics[scale=0.07]{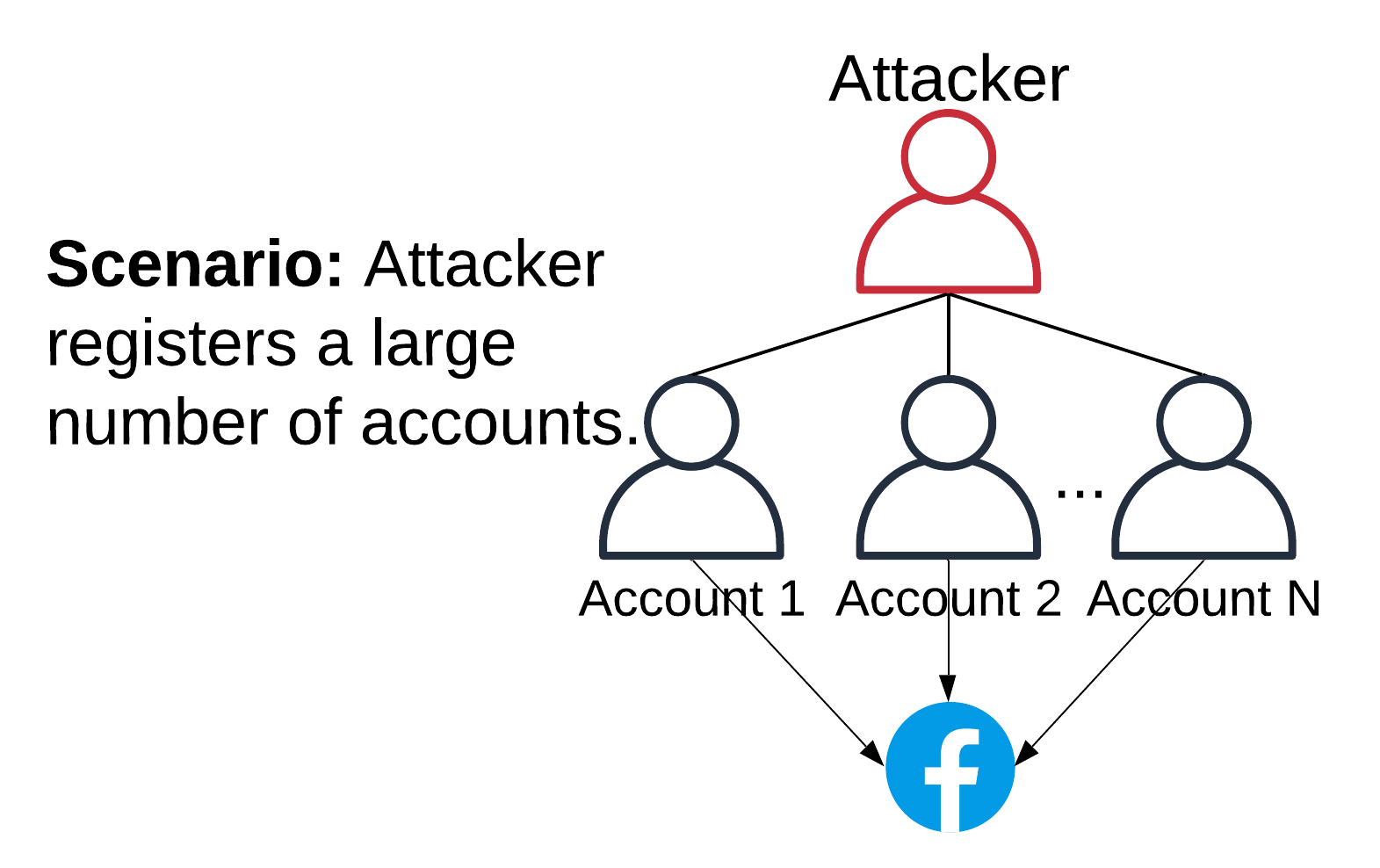} }}%
    \qquad
    \subfloat[Sockpuppet]{{\includegraphics[scale=0.07]{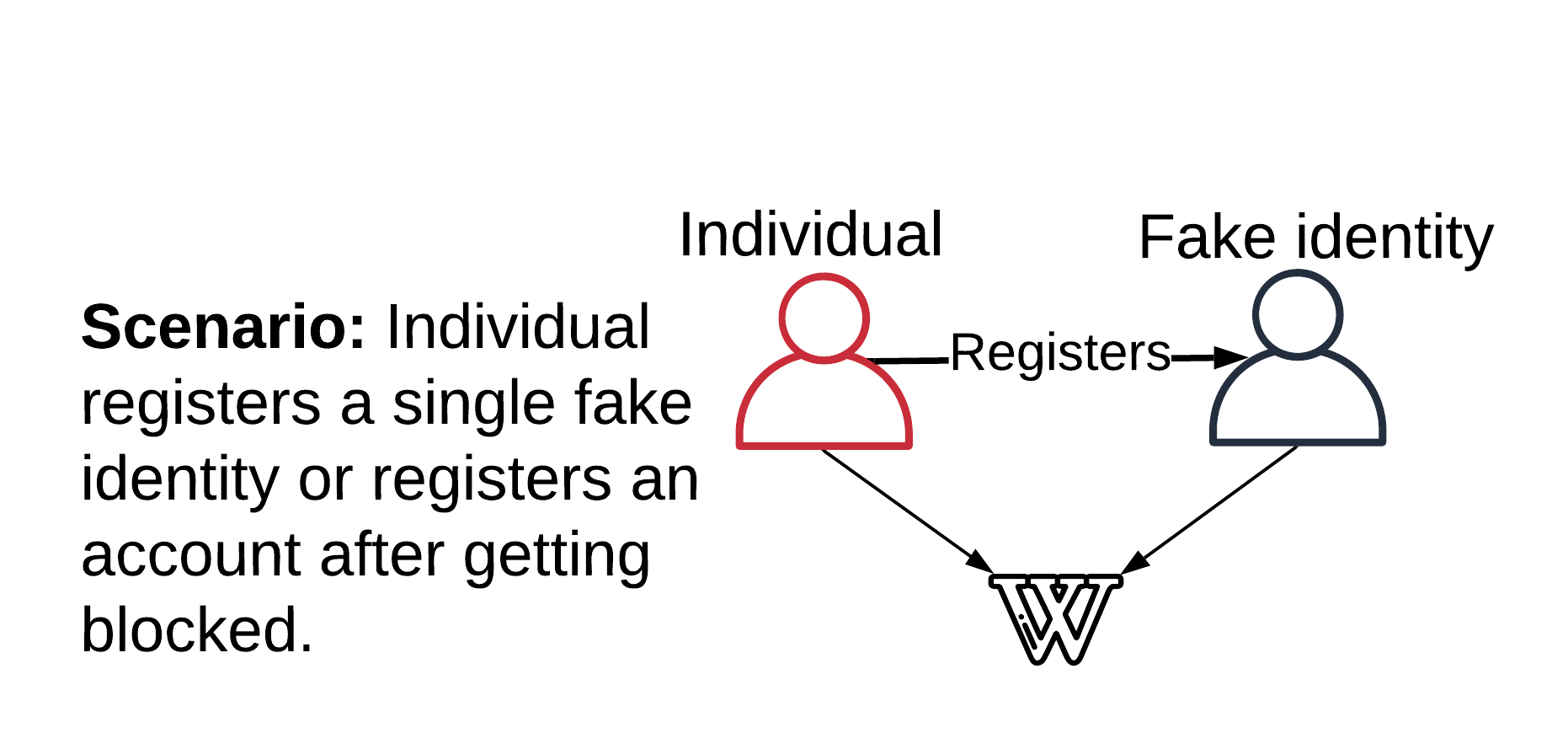} }}%
    \qquad
    \subfloat[Social botnet]{{\includegraphics[scale=0.07]{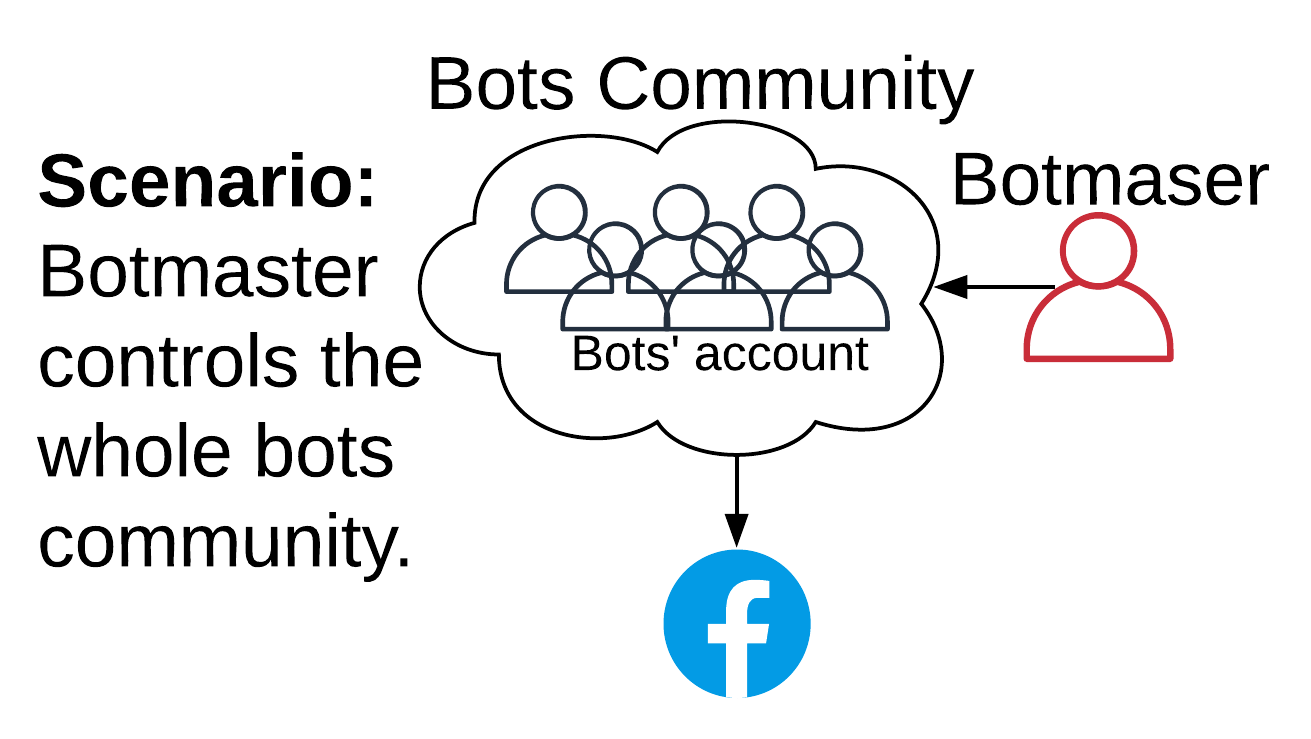} }}%
    \caption{An example of fake profile attacks}%
    \label{Fake_profile}%
\end{figure}

\paragraph{Definition:}
A sockpuppet is defined as a single fake online identity which has been created by an individual (or puppetmaster) \cite{bu2013sock}\cite{Solorio2013}\cite{Wang2019}, as shown in Figure \ref{Fake_profile}b. Other researchers have defined sockpuppets as the creation of new malicious accounts after the previous accounts being detected and blocked \cite{Solorio2013}. The main objective of creating sockpuppet accounts is to unfairly support a user's point of view on a topic such as movie reviews and business promotions \cite{Maity2017}. Sockpuppet accounts can also be used to manipulate online polls. For example, these accounts can be used to submit multiple votes suggesting the representation of a majority opinion or sideline opposition voice.
\paragraph{Use Case:} 
A user registers an account to publish a fake article or vandalize existing articles in Wikipedia. Wikipedia administrators ban the account from publishing and editing articles. The user then registers a new account to circumvent the ban.

\subsubsection{Social Botnet}\label{socialbot}
\paragraph{Definition:}
A botnet is a collection of social bots (malicious users). A human botmaster controls the social bots. He or she constructs and controls the whole bots community \cite{Chu2012}\cite{Zhang2018b}, as shown in Figure \ref{Fake_profile}c. In addition, social botnets communicate with social media users and then establish communications with the botmaster. Social botnets are also fairly easy to create because the open APIs are published by social media platforms.
\paragraph{Use Case:}
An adversary (botmaster) constructs and controls a group of social bots (malicious users) to produce deceptive content in social media platforms. A social botnet performs malicious actions such as publishing spam content, spreading malware and phishing online social media \cite{Chu2012}. Zhang et al. \cite{Zhang2018b} paid \$57 to purchase 1000 bot accounts on Twitter.

\subsection{Identity Theft} \label{identitytheft}
Identity theft happens when personal information is stolen in social media platforms \cite{Wang2006a}. Identity theft exploits the vulnerabilities of the victim's personal account(s) or device(s). The adversary might gain a partial (compromised account) or full (hijacked account) control of the victim's account, as shown in Figure \ref{identity_theft}a. We discuss these two subcategories in the following subsections.
\subsubsection{Compromised Account} \label{compacc}
\paragraph{Definition:}
A compromised account is an existing and legitimate account to which an unauthorized person has gained access or control \cite{Egele}. The victim can still access his/her account when the account is compromised, as shown in Figure \ref{identity_theft}b. There are several ways to compromise accounts; for instance, by exploiting vulnerabilities of the victim's account or by using a phishing scam to steal the victim's login credentials. This exploitation causes the victim's friends to click on malicious links because they believe that the origin of the links is legal \cite{Seyler2018}. The compromised account is hard to detect since its profiles exhibit the original user's information \cite{Ruan2016}. The account holder's friends are unaware of the account having been compromised.  Consequently, attackers operate the account, publish posts, send friend requests, etc \cite{kaur2018authcom}.
\paragraph{Use Case:}
Accounts are compromised in different ways. An attacker may take advantage of weak passwords to get control of an account. Alternatively, an attacker may use phishing sites to ask a user to provide his account details. In addition, an attacker may take advantage of a third-party application. For example, an account gets compromised by Java applet on a website or by downloading a plugin that may be malicious.
\subsubsection{Hijacked Account} \label{hijacc}
\begin{figure}%
    \centering
    \subfloat[Identity theft]{{\includegraphics[scale=0.08]{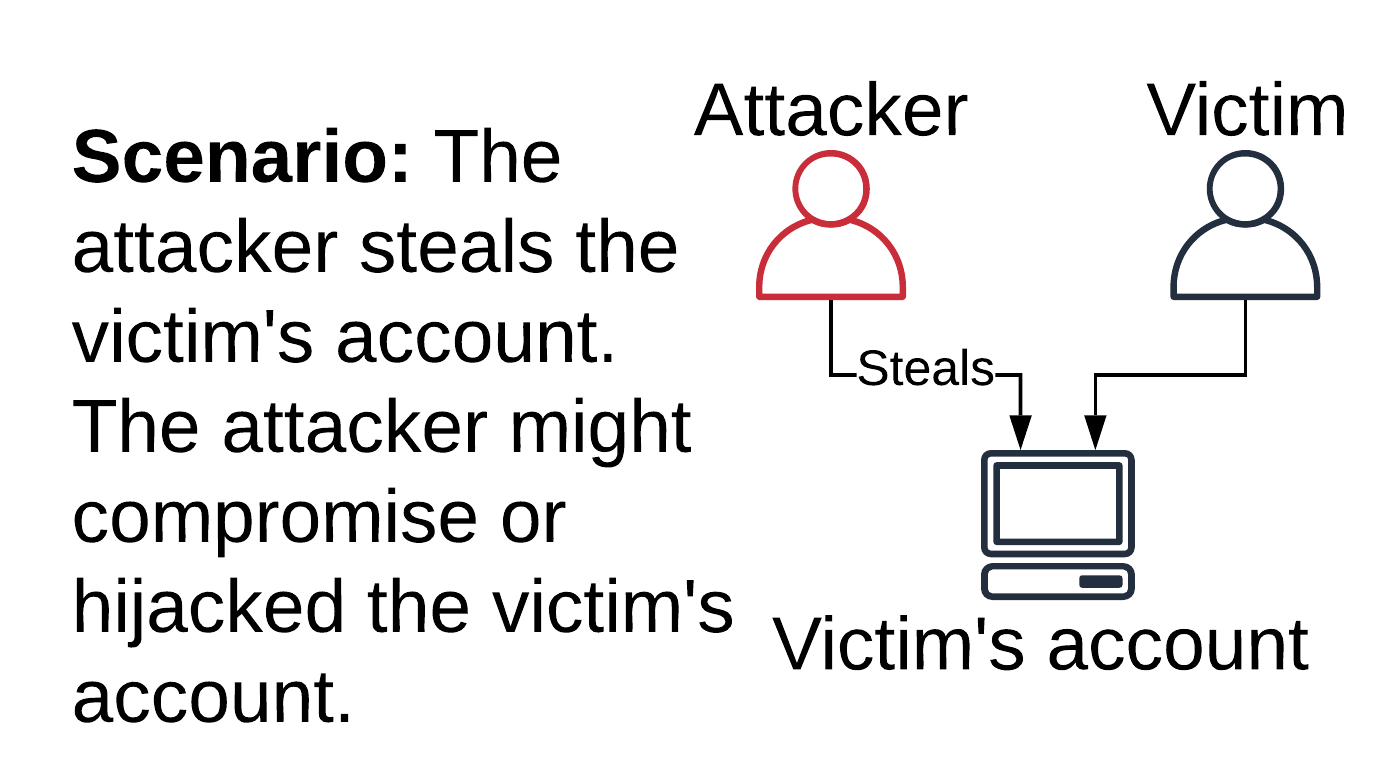} }}%
    \qquad
    \subfloat[Compromised account]{{\includegraphics[scale=0.08]{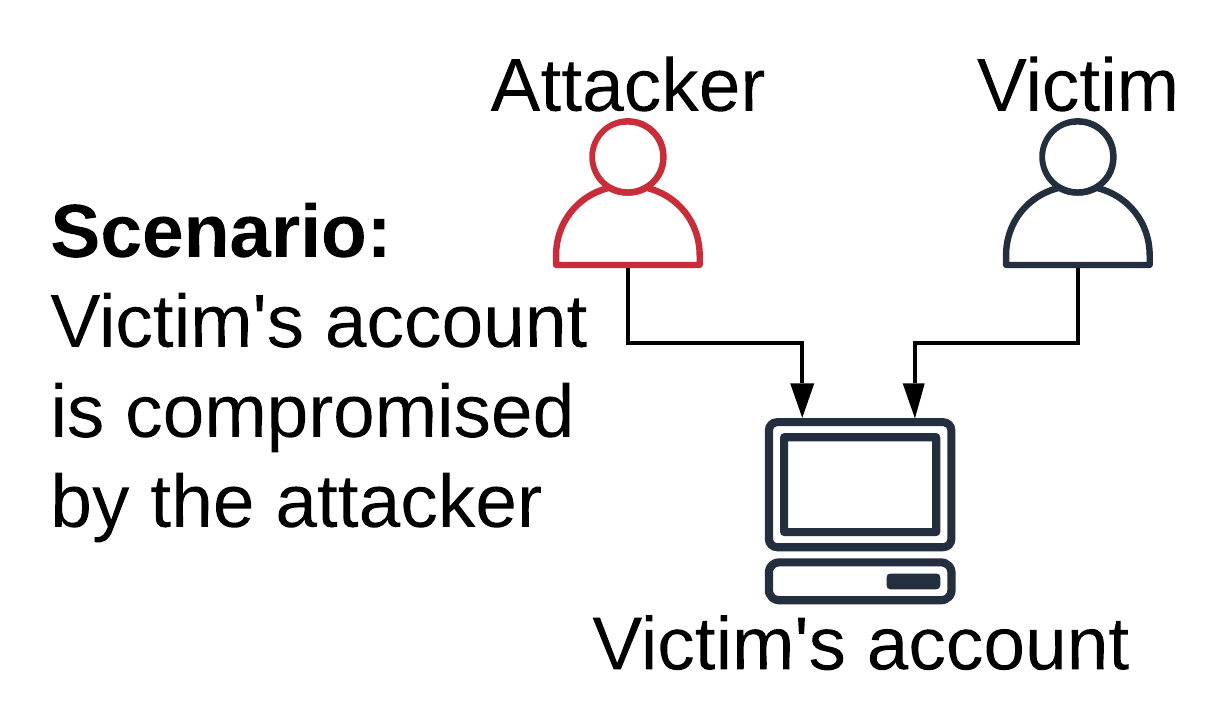} }}%
    \qquad
    \subfloat[Hijacked account]{{\includegraphics[scale=0.08]{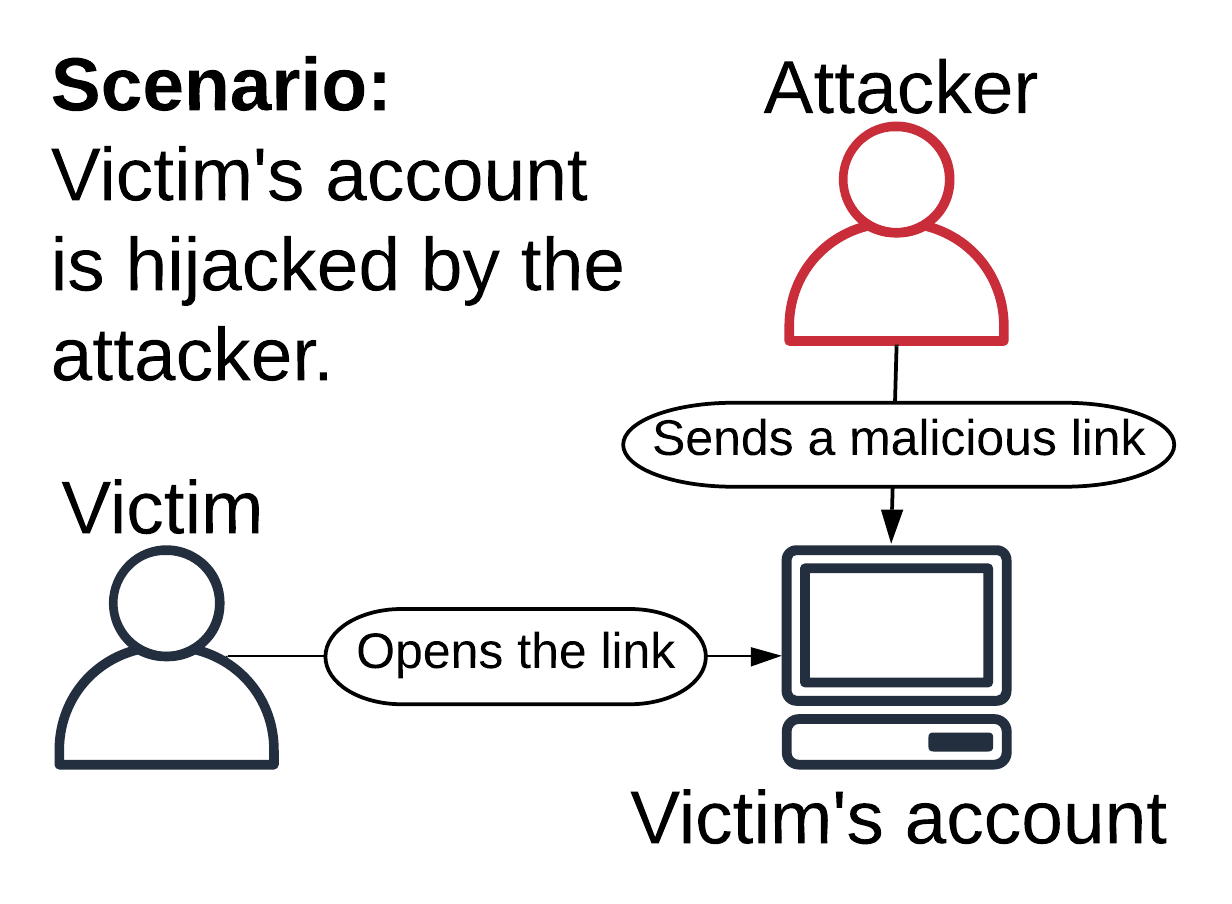} }}%
    \caption{Examples of identity theft attacks}%
    \label{identity_theft}%
\end{figure}

\paragraph{Definition:}
A hijacked account is similar to a compromised account, except that a victim loses full control of his/her account to an attacker in the hijacked account \cite{Bahri2018}, as shown in Figure \ref{identity_theft}c.
\paragraph{Use Case:}
Assume that an attacker wants to hijack a victim's account. The attacker uses a malicious program which can steal the victim's login credential saved in the victim's Web browser or computer. The attacker uses the stolen credential to login the victim's account and changes the victim's password. As a result, the attacker will have full control of the account until the victim restores his/her account.

\subsection{Identity Cloning} \label{identityclonin}
Identity cloning is to create a fake profile using stolen identity information. In particular, an adversary imitates a victim's account values to collect the private information of the victim's friends. In addition, the adversary increases the trust level with the victim's friends for future fraud. Identity cloning attacks can be divided into single-site and cross-site profile cloning \cite{Bilge2009}\cite{Jin2011}. It has been reported that tens of thousands of accounts were created using stolen real personal information. Jessica Rychly was one of the victims of identity cloning. Her Twitter account information
(her profile photo, location and biographical information) was cloned onto a fake account that retweeted graphic pornography and cryptocurrency advertisements\footnote{https://www.nytimes.com/interactive/2018/01/27/technology/social-media-bots.html}.

\subsubsection{Single-Site Identity Cloning} \label{sinsite}
\paragraph{Definition:}
Single-site identity cloning refers to an attacker creating a cloned profile of the victim in the same social media platform \cite{Bilge2009}.
\paragraph{Use Case:} 
An adversary first exploits a victim's public personal information on social media such as occupation, name, friends list and location \cite{Jin2011}. The adversary then creates a similar or identical profile of the identity of the victim in the same social media, as shown in Figure \ref{identity_cloning}a. The adversary then sends friendship requests to the victim's friends. The adversary waits until a few of the victim's friends (blue) accept the friendship request. Next, the adversary obtains access to those friends' profiles.
\subsubsection{Cross-Site Identity Cloning} \label{crosssite}
\begin{figure}%
    \centering
    \subfloat[Single-site identity cloning]{{\includegraphics[scale=0.10]{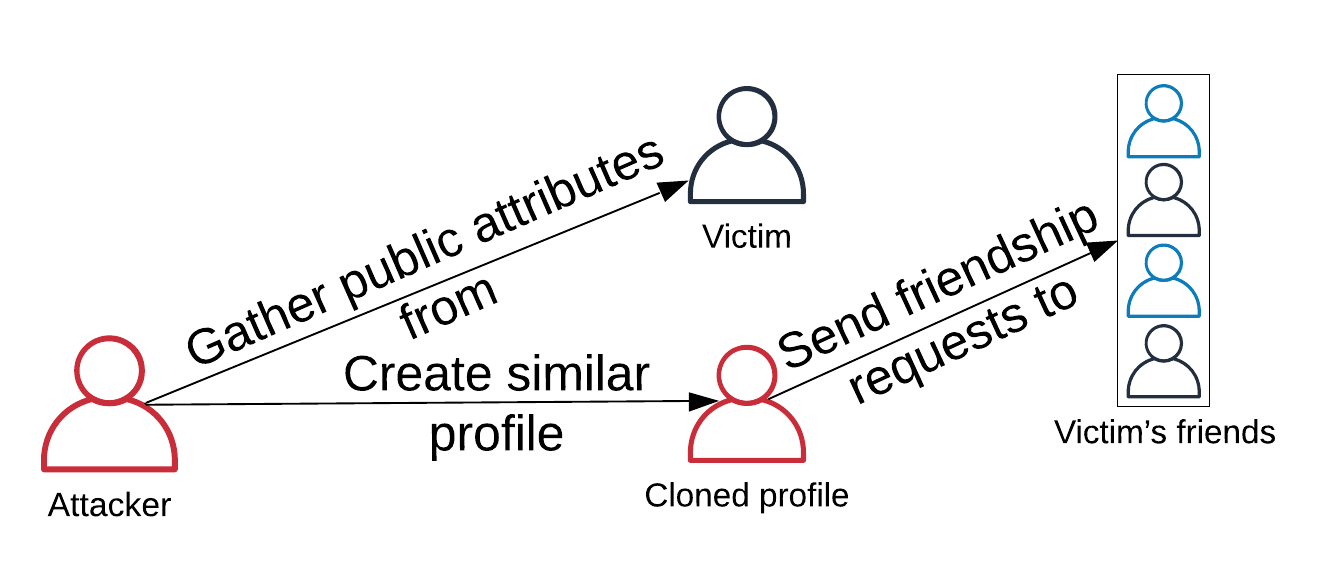} }}%
    \qquad
    \subfloat[Cross-site identity cloning]{{\includegraphics[scale=0.10]{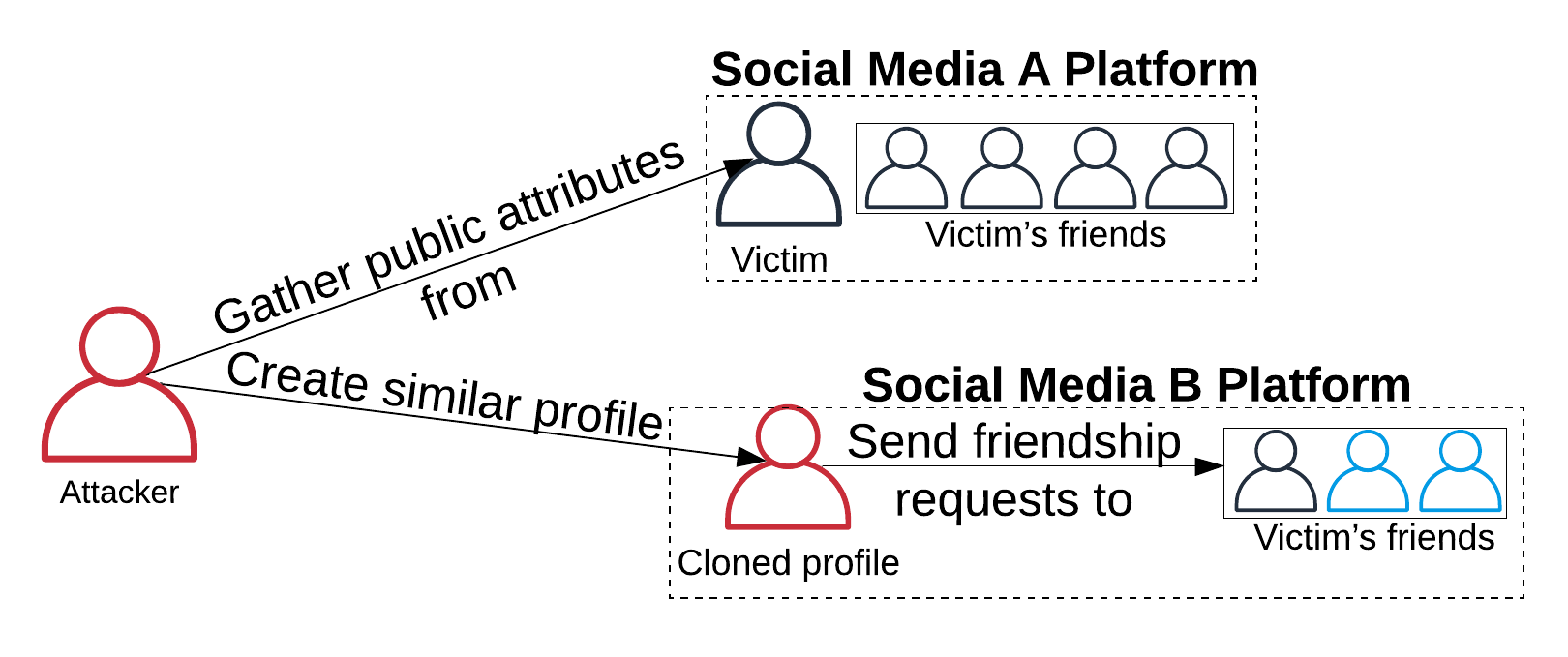} }}%
    \caption{Examples of identity cloning attacks}%
    \label{identity_cloning}%
\end{figure}

\paragraph{Definition:}
Cross-site identity cloning refers to an attacker cloning the profile of a victim from a social media platform to other social media platforms where the victim has not yet registered \cite{Bilge2009}.
\paragraph{Use Case:} 
An adversary first gathers a victim's public personal information on social media such as name, friends list, occupation and location \cite{Jin2011}. The adversary creates a similar or identical profile of the victim’s identity in another social media, as shown in Figure \ref{identity_cloning}b. The adversary then sends friendship requests to the victim’s friend list. The adversary waits until the friendship requests are accepted. A few of the victim's friends (blue) accept the friendship request. The victim’s friends cannot clearly differentiate between the cloned and real identities because the private attributes of the real profile are not displayed in the cloned profile.  Therefore, the victim’s friends are more likely to accept the friendship request. Therefore, the adversary obtains access to the friends' profiles.  

\subsection{Comparison of Identity Deception Attacks} 
The aforementioned identity deception attacks are representative of the most common identity deception attacks on social media platforms reported to date. Table \ref{Comp_identity_attacks} summarizes the major features (i.e., purpose, procedure, outcome, and target networks) of the social media identity deception attack categories. 
These identity deception attacks are deeply intertwined. An attacker can initiate one type of attack in the form of another type of attack. For example, the attacker can initiate a Sybil attack in the form of social botnet. Some types of identity deception attacks may require the assistance of another type of attacks. For instance, the Sybil attack can take advantage of the sockpuppets' accounts to launch deceptive attacks. 

\begin{table*}[]
\caption{Comparison of most popular identity deception attacks on social media}
\label{Comp_identity_attacks}
\resizebox{\textwidth}{!}{%
\begin{tabular}{p{0.20\textwidth}p{0.25\textwidth}p{0.25\textwidth}p{0.26\textwidth}p{0.21\textwidth}}
\toprule
\textbf{Attack}            & \textbf{Purpose} & \textbf{Procedure} & \textbf{Outcome} & \textbf{Target Network} \\
\toprule
\textbf{Fake Profile}      &         &           &                &                \\ 
\hspace{3mm}Sybil attack & To subvert or pollute the social media reputation with large number of fake identities.  & The adversary registers a large number of fake identities.   & Acquiring a disproportionate level of control over the social media and increasing its power within the social media.        &     Peer-to-peer networks, Microblogging and Social media sites.          \\ [-0.85em] \\

\hspace{3mm}Sockpuppet& To support point view or opinion or to circumvent restrictions. & The individual registers a single fake identity. & Gaining unfair support to a point of view or circumventing restrictions.      &    Microblogging, Collaborative projects and Blogs.            \\ [-0.85em] \\
\hspace{3mm}Social botnet&To automate the scope of an attack.& A computer algorithm is used in order to produce malicious content and communicate with social media users.&The scope of the attack is expanded. &  All types of social media.              \\
\toprule
\textbf{Identity Theft}    &         &           &                &                \\
\hspace{3mm}Compromised &To defame or deceive a social media user, to abuse legitimate accounts and to spread false information.         &  Social-engineering attacks such as phishing are used to obtain passwords or depending on the fact that most people do not change their passwords across most social media platforms.         & The original user does not have full control over his/her account, i.e., account owner and adversary access the same account.               &            All types of social media.  \\ [-0.85em] \\
\hspace{3mm}Hijacked                  &     As above    &     As above      &       The original user cannot access his/her account.         &       As above         \\
\toprule
\textbf{Identity Cloning} &         &           &                &                \\

\hspace{3mm}Single-site or cross-site & To steal someone's personal information and to deceive or defame a person. &

  A cloned profile of the victim in the \textit{\textbf{same}} or \textit{\textbf{cross}} social media platforms is created. \newline Friendship requests are sent to the victim's friends. \newline  A friendship request is accepted based on the trust between them.
 & A new network of the victim is established where the adversary able to access the private information shared in the network. &     All types of social media.            \\

\toprule
\end{tabular}%
}
\end{table*}

\section{Detection Techniques of Social Media Identity Deception}
In this section, we analytically review the existing detection techniques for each type of identity deception attack defined in Section 3. 
\subsection{Fake Profile Detection Techniques}
Many research efforts have been made to detect fake profiles in social media platforms. Researchers have proposed various detection techniques to minimize fake profiles in social media platforms. 
\subsubsection{Sybil Attacks Detection Techniques}
We categorize the Sybil detection techniques into three classes: \textit{graph-based techniques, machine learning-based techniques and key management-based techniques} in terms of their adopted technologies.
\paragraph{Graph-Based Techniques}
These techniques employ social structures which are represented by social graphs. Consider a graph \(G = (V, E)\), where a user is represented by a node  \(v\in V\) and  a mutual relationship between \(u\) and \(v\) is represented by an edge \((u, v)\in\ E\). For example, edge \((u, v)\) indicates that \(u\) and \(v\) follow each other in Twitter. Graph-based techniques use social network information and assume that Sybil accounts cannot build a relationship with legitimate accounts quickly. Therefore, the number of edges between the Sybil account and the legitimate account is restricted in social networks which usually have a high level of trust \cite{Al-Qurishi2018}\cite{AbedelazizMohaisen2010}. Sybil detection techniques mostly base on Random Walk (RW) techniques and Loopy Belief Propagation (LBP) techniques \cite{yu2008sybillimit}\cite{yu2006sybilguard}.
RW randomly chooses and navigates to one of the neighbour nodes through the edges connected to the initial node. This process is iterated until all the nodes in the graph have been visited \cite{spitzer2013principles}. 
LBP is a dynamic programming approach to solve conditional probability queries in graphs. For each unobserved node, it computes the marginal distribution, conditional on any observed nodes. These techniques iteratively propagate the label information in social graphs to predict the account labels \cite{ihler2005loopy}.

Jia et al. \cite{Jia2017} proposed SybilWalk for Sybil identities detection. SybilWalk bases on the RW. It overcomes  the drawbacks of the existing RW-based methodologies by taking advantage of their desired features. The main idea behind SybilWalk is to capture the structural gap between Sybil nodes and benign nodes by label-augmented social network-based RW. First, a label-augmented social network was implemented to combine both labelled Sybils and labelled benign nodes in the training dataset. Two nodes are added to the existing social network:
benign \(l_b\) node and Sybil \(l_s\) node. 
Next, badness scores are defined using the RW as follows:
\begin{equation}
\centering
p_u^{(t)} = \sum_{ v \in \Gamma _u } \frac{w_u{_{v}}}{d_u}  p_v^{(t-1)}
\end{equation}
where \(p_u\) is the badness score of a user \(u\), \(\Gamma _u\) is the set of \(u\) neighbors, \(d_u\) is the weighted degree of \(u\) and \(w_u{_{v}}\) is the weight of edge \((u,v)\).
The badness score of each user node is computed by an iterative approach. A high probability of being Sybil is indicated by a higher badness score. 

Zhang et al. \cite{Zhang2018} proposed algorithms to detect Sybil accounts. They implemented a social and activity network, which is a two-layer hyper-graph, to integrate users' historical activities and friendships. The authors designed a Sybil social and activity network to detect Sybil via an RW based technique on the social and activity networks. The trust score of each node was evaluated by applying the RW technique. The motivation behind the proposed RW technique is the mutual reinforcement relationship between activities and users. The trustworthiness of the user can be verified by a high trust value of the activity. Based on that, the social and activity network is decomposed into three subgraphs: activity-following graph, user-activity graph and friendship graph. An RW was designed to propagate trust independently for each subgraph. Finally, the mutual reinforcement relationship between activities and users is captured by the combined RW technique. 

Wang et al. \cite{Wang2018b} proposed an approach to combine LBP-based approaches and RW-based approaches into a local rule-based framework to detect Sybils called SybilSCAR. First, prior knowledge of all nodes is allocated by RW and LBP. The prior knowledge is then propagated through the social network in order to get the posterior knowledge. The authors iteratively applied the local rules of posterior knowledge for each node. A local rule updates the posterior knowledge of a node by integrating the influences of its neighbours with its prior knowledge. The new local rule incorporates the advantages of both LBP-based and RW-based local rules. The main purpose is to employ the multiplicativeness. This means that LBP-based local rule is robust to noisy labels. It is also space and time inefficient. SybilSCAR iteratively applies the new local rule for each node to calculate the posterior probabilities.
SybilSCAR first utilizes labelled Sybils  and labelled benign nodes.
A prior probability to become Sybil for all nodes is then assigned.

Gao et al. \cite{Gao2018} proposed a defence-in-depth framework to detect Sybil accounts called SYBILFUSE. Unlike the existing works on Sybil detection that base on oversimplified assumptions about network structure, SYBILFUSE adopts a collective classification scheme in order to relax these limitations. First, SYBILFUSE samples training data. It also leverages local attributes for local classifiers training. Each node trust score is predicted to indicate node probability of being benign. Each edge trust score is also predicted to indicate the edge's probability of being a non-attack edge. These local scores are then amalgamated with the global structure by weighted trust score propagation. Furthermore, edge trust scores are leveraged to impose unequal weights for the propagation. As a result, the attack edges impact will be reduced. The propagated node scores are used for Sybil ranking and prediction. Weighted RW and weighted LBP are used. Equation \ref{eq:1} is used to update the weighted RW and Equation \ref{eq:2} is used to update the weighted LBP. 
\begin{equation}\label{eq:1}S^{(i)}(v) = \sum_{(u,v)\epsilon E} S^{(i-1)}(u)\frac{S_{u,v}}{\sum(u,w)\epsilon E^{S_{u,w}}} \end{equation}
\begin{equation}\label{eq:2} S_{V}^{F} = \frac{bel_{v}(X_{v}=1)}{bel_{v}(X_{v}=1)+bel_{v}(X_{v}=-1)} \end{equation} 
where \(S^{(i-1)}(u)\) is the score of node \(v\) after \(i\)-th power iteration, \(u\) and \(v\) represent the node, \(w\) represents an unfamiliar connection, \(X\) is a binary random variable and \(bel_{v}(X_{v}=1)\) is the belief score of node \(v\) with label \(X_v = x_v\). 

Yang et al.  \cite{Yang2018}  proposed a Sybil defence technique that models user friend-request as a signed graph and performed Sybil detection by embedding it into a low-dimensional space. This technique has three stages: 1) finding an embedding that is a local optimum through minimizing the objective function using stochastic gradient descent, 2) enabling the embedding moving towards the global location as one node and 3) identifying the Sybil region based on the embedding result.

A downside of RW-based techniques is that they only leverage either labelled honest users or labelled Sybils in the training dataset. This limits their detection accuracy. RW-based techniques are also not robust to noises in the training dataset 
\cite{Jia2017}\cite{Wang2018b}. LBP-based techniques are unscalable to adapt to real-world social networks. Within the networks with hundreds of millions of users and edges, LBP-based techniques cannot guarantee convergence \cite{Wang2018b}. 
\paragraph{Machine Learning-Based Techniques} 
These techniques formulate historical user behaviours by tracking their activities on the social media platforms during the period of their registration. A user’s actions are contained in connections between the user profile and the other elements of a social media platform. As a result, the user’s actions can be easily tracked. An example of the connections is the relationships originating from the user’s profile. In addition, communications between two users are considered to be a connection \cite{Al-Qurishi2018ba}.  Researchers leveraged machine learning-based techniques to search for users who break the rules. They compared the expected users' activity patterns and observe the infractions. An account is represented by a set of features extracted from account profile information and accounts' social graphs. A training dataset consists of labelled honest accounts and labelled Sybils. It is fed into a binary classifier, e.g., support vector machine (SVM). Finally, the classifier is applied to predict labels for the remaining accounts.

Cai and Jermaine \cite{Cai2012} proposed a latent community (LC) model to detect Sybil attacks in a collaborative network. LC is an associated learning approach and statistical model. The nodes in a graph are partitioned into closely linked communities that have strong internal connections. Interconnections of the node in a graph are statistically described. The LC positions are generated based on two distributions: 1) Gaussian distribution, which positions the legitimate communities close to the centre of the space and 2) spherical distribution of adversary, which surrounds the Gaussian distribution.  Malicious nodes are established to probably belong to the mixture component linked with the adversaries. 

Mulamba et al. \cite{Magno2014} classified users into honest and malicious types for Sybil detection. This research focuses on a classification mechanism that leverages the social network graph topologically. A new set of features based on centrality metric was defined. The features were obtained from social graph structure. It also describes the node importance in a graph. The main concept behind these features is to propose features that an adversary cannot exploit to avoid detection. The authors used a feature vector to train three models: k-nearest neighbors, random forest and Adaboost. They employed the Recursive Feature Elimination \cite{Iguyon2003} to select the features. 

Al‐Qurishi et al. \cite{Al-Qurishi2018} developed a new defence system to minimize Sybil attacks.
The idea is to conspire a dataset  \(D={L_D\cup U_D}\), where \(L_D\)  is a small set of labelled data and \(U_D\)  is a large set of unlabelled data. The problem is transformed into the form of a semi‐supervised transductive learning technique. The proposed technique relies on RW-based label propagation. Labels of unlabeled data points in \(U_D\) are predicted. The energy function \(E(V_{ij})\) was computed to capture the two pairs of nodes in the graph, as represented in Equation \ref{eq:3111}, where \(L_W\)  is the local weight,  \(W_{ij}\) is the global weight and \(i\) and \(j\) are two nodes in the graph. An RW with absorbing states is adopted. The interaction graph was used to label the propagation process. The extracted features are categorized into user profile, content level and interaction network. Three types of interaction graphs are built using a Twitter dataset, including retweet graphs, reply graphs and mention graphs. Each graph has a set of targeted accounts and reveals the interactions of each account in the set.
\begin{equation}\label{eq:3111} E(V_{ij})= \frac{L_{w(V_j)}\times W_{ij}}{\sum_{k\in Adj}L_{W(V_{k})}\times W_{jk}}  \end{equation} 

Al‐Qurishi et al. \cite{Al-Qurishi2018ba} proposed a deep learning model that detects Sybil attacks on Twitter. The proposed system consists of three modules: 1) data harvesting module, 2) feature extracting mechanism and 3) deep-regression model. In the data harvesting module, the Twitter API was used for extracting various tweets of the user's account. 
In the feature extracting mechanism, the harvested data are divided into three categories. First, profile-based features are taken from a user profile. The content-based features are then divided into four features based on temporal, emotion, quality  and topic. Next, in graph-based features, the proposed system tracks three separate matrices: for addressing, for co-appearing hash signs (\#), and  for republishing. The deep-regression model, which is a feed-forward neural network, receives features vectored from the upper component. This model classifies the features vector into categorical features and continuous features. The categorical features consist of strings and binary values. The continuous features consist of numerical values. 

A fundamental limitation of these machine learning-based techniques is that the adversaries can manipulate the profile information to mimic honest accounts. These adversaries can therefore bypass the detection. In addition, machine learning techniques rely on relatively larger datasets to train an accurate classifier. 

\paragraph{Key Management-Based Techniques}
These refer to management of cryptographic keys in a cryptosystem. The main idea of such techniques is that a suspected node can generate private and public keys for identity protection and prevent them from being stolen by an attacker. 

Al‐Qurishi et al. \cite{Al-Qurishi2018a} proposed an m-resilient authenticated centralized key management protocol. The proposed protocol is able to curb Sybil attacks and produce secure communication between social networks users.
The protocol presumes that online social networks have a cryptographic private key generator. Each time an individual registers a new account, he or she chooses a non-existing username and provides it to the private key generator. The latter applies a special hash function on the unique identifier (ID) to generate a cryptographic ID. A private secret is generated by the private key generator using the cryptographic ID. It is originated from the private key generator’s master secret merged with pairing-based cryptography. An account can dynamically generate a random session key using its neighbour nodes using only one secret. Therefore, the proposed protocol considerably reduces key storage. A defect of this technique is that it provides inappropriate key storage. Keys cannot be stored with the data on a database or server, since any exfiltration of the data might compromise the key \cite{das2010g}\cite{dong2011shared}.

Table 1 in Appendix presents an overview and comparison of the aforementioned papers based on the goals, techniques, features  and datasets. 

\subsubsection{Sockpuppet Detection Techniques}
The puppetmasters create more than one accounts to avoid getting blocked, to support their propaganda and  manipulate social media content. Sockpuppet detection techniques aim to detect whether the same user controls more than one account. 
We categorized sockpuppet detection techniques into three classes: \textit{verbal behaviour analysis, non-verbal behaviour analysis and similar-orientation network.}  
\paragraph{Verbal Behaviour Analysis} It relies on the identification of authorship attribution (AA). AA refers to the process of text writer identification, which captures the writing style of a writer \cite{coyotl2006authorship}.

Solorio et al. \cite{Solorio2013} presented an automated sockpuppet detection approach in Wikipedia based on machine learning. The sockpuppets problem is solved based on the AA viewpoint. The proposed approach analyses the written document to predict the real author. It links the sockpuppet accounts to the corresponding puppeteer based on the writers’ unique style. The proposed approach consists of two steps. 1) It gathers predictions from the classifier on each comment. 2) It combines them in a voting schema for final decision. 239 features were utilized, which capture grammatical, stylistic and formatting preferences of the writer. An SVM was used to perform classification. The limitation of this approach is its high computational cost. 
It requires a time complexity of $O((N  R)^2)$ to test all the users in a datasets. Testing every new user against all the existing users in a dataset will need a time complexity of $O(N \times R)$, where $R$ is the number of revisions made by a user.

Zheng et al. \cite{Zheng2011} proposed two methods of detecting a sockpuppet pair in online forums. A system that is able to gather posts from 11 popular Hong Kong forums was implemented. The system has four stages: Crawlers, Database, Data-mining engine and Statistical engine. The crawler automatically collects the posts. The database is responsible for storing the collected posts. Data-mining and statistical engines are accountable for the posts analysis. The first algorithm detects sockpuppet in one forum. It is a simple method for sockpuppet pairs detection without considering the posts' contents. It bases on the total (weighted) amount of topics published by one account and the relative number of replies by the other account concerning all his replies. The second algorithm detects sockpuppet in two different forums.  It creates a keyword-based profile based on the corresponding posts. It extracts the keywords from the two different forums and then compares the similarity of the keywords of each forum to decide whether it is a sockpuppet pair or not. 

Crabb et al. \cite{Stephanidis2015} employed a character n-gram approach for sockpuppet detection. A naive Bayes classifier was constructed using normalized frequencies of parsed character bigrams to contrast the use of  writer bigram. A computational algorithm was used for non-UTF-8 characters detection. The non-UTF-8 characters that produce highly probable uninformative features
are removed from Twitter datasets. Next, n-gram features are extracted and decomposed into their component character bigrams. Each bigram frequency was calculated at the tweet level for values' normalization. A feature matrix that displays the frequencies of each bigram normalized is the output of n-gram features extraction. Finally, a naive Bayes classifier was implemented to classify the sockpuppets.

Johansson et al. \cite{Johansson2014} proposed four various types of matching methods to detect aliases. The proposed methods base on string, stylometric, time profile and social network. String-based matching depends on the names of aliases. First, a Jaro-Winkler distance measure \cite{winkler1990string} was implemented. 
Time-based matching depends on the written posts. Time profiles are created on the time relative distribution on the day when the tweet is written. Each day time is divided into equal intervals representing one hour. Euclidean distance was utilized to calculate the distance between two time profiles vectors.
If the distance among the time profiles for two aliases is small, the two aliases are more likely to belong to the same user. Stylometric matching depends on written posts. A ``writeprint'' was constructed in order to analyse the alias’s writing style.  Social network-based matching depends on thread or friend information. The social network of two aliases is mapped and compared in order to reveal if those aliases are similar. A limitation of this technique is that the  temporal features that the technique rely on  the total number of posts available for experiments.

A common pitfall of verbal behaviour analysis is that this type of techniques mostly suffers from high computational complexity, the scalability of which is questionable. 


Table 2 in Appendix presents an overview and comparison of the aforementioned papers based on the goals, techniques, features and datasets. 

\paragraph{Non-Verbal Behaviour Analysis} This bases on identifying a sockpuppet account by extracting features that capture user activity or movement.

Tsikerdekis et al. \cite{Tsikerdekis2014} proposed a multiple account identity deception detection technique based on non-verbal behaviours. Wikipedia was used as an example of social media. The concept of namespaces is operated on Wikipedia. Each namespace is intended to incorporate a particular sort of content. An example of a non-verbal behaviour is the time taken between each revision. Time-independent and time-dependent variables are utilized to represent user behaviours. All user activities in the first 30 days were obtained. Standardized difference between sockpuppets and legitimate users was computed. Finally, several models were experimented: Adaboost, SVM and random forest. A shortfall of the proposed detection technique is that the time window size for observing the behaviours of the newly registered users has a significant impact on the effectiveness of the proposed technique. It can also impact the efficiency if the window size is too large, given that more data will be needed to be examined by the proposed detection technique.

Kumar et al. \cite{Kumar2017} studied sockpuppetry characterized in nine different online discussion communities. The authors stated that sockpuppets and legitimate accounts differed in terms of their writing behaviours and interacting with other sockpuppets. First, sockpuppetry behaviours were studied in terms of how the sockpuppets are created, their language and social networks. The proposed method is driven by the manner in which Wikipedia administrators identify sockpuppets. The authors distinguished sockpuppets from legitimate accounts and identified pairs of sockpuppets in the communities by using three types of features: activity, community and post features. 

Yamak et al. \cite{Yamak2018} proposed a methodology to detect and group sockpuppets, called ``SocksCatch''. SocksCatch has three stages: data collection and selection, sockpuppet accounts detection and sockpuppet accounts grouping. 
In the sockpuppet accounts detection phase, a proposed features' set was computed for each chosen user. Also, the computed features were evaluated on a trained and tested machine learning algorithm for sockpuppet detection. The proposed feature subtracts the time of the performed action from the account's joining time. Sockpuppet accounts grouping phase bases on accounts’ actions and attributes. Each detected sockpuppet account is related to the pages where it performs actions in order to form the action graph. SocksCatch was applied to real-world data from a collaborative project which is English Wikipedia. The authors evaluated SocksCatch's efficiency based on six machine learning models: SVM, random forest, naive Bayes, Bayesian network, k-nearest neighbour and Adaboost. 

Xiao et al. \cite{Xiao2015} proposed a time-sensitive and scalable machine learning technique to detect fake users created. Unlike the existing works only focusing on purpose-built fake users prediction techniques for each user, this study classifies fake users to determine whether those users are malicious or legitimate. The features available at registration time or shortly thereafter were used. Three approaches were employed: random forest, SVM and logistic regression. The proposed approach consists of three parts: cluster builder, profile featurizer and account scorer. First, the raw list of users was clustered along with their raw features. Next, each cluster was converted into a single numerical vector. 

The limitations of the non-verbal behaviour analysis include: 1) it is only designed for limited social media platforms and 2) it can only detect a sockpuppet account at one time, which is less effective in a real-world environment.

Table 3 in Appendix presents an overview and comparison of the aforementioned papers based on the goals, techniques, features and datasets.

\paragraph{Similar-Orientation Network} Similar-orientation network bases on evaluating the similarity of sentiment orientations among user account pairs to construct a similar-orientation network \cite{Wang2019}. The relationship between two users is constructed if the two users have a comparable view of the most topics and comparable writing styles \cite{Li2019}\cite{Wang2019}\cite{Makinouchi2018}.

Liu et al. \cite{Liu2016} proposed a solution for detecting multiple accounts which are managed by an organization or person as a sockpuppet gang. The authors solved this problem from a data mining viewpoint. First, the sentiment orientation of the user's comments was analysed based on emotional phrases. Next, if the similar orientations to most topics exist between user account pairs, a similar orientation network was built. In the similar-orientation network, each node indicates a user account and an edge exists only if sentiment orientations to most topics have existed between two user accounts. The authors proposed multiple RWs on a similar orientation network to re-measure the weight of each edge iteratively based on analysing user account behaviours. 

Wang et al. \cite{Wang2019} proposed a structure-based online approach to detect sockpuppet accounts.  First, the difference between sockpuppet pairs and sockpuppet legitimate pairs was analysed based on social networks. The proposed approach constructs the social network graph for each user based on the analysis. The social network graph consists of an interest and interaction network graph. The similarity of the social networks between users was then analysed by subgraph similarity matching.

Li et al. \cite{Li2019} proposed a propagation tree for sockpuppet detection. This study illustrates that the sockpuppet and the legitimate account have unusual patterns in the propagation trees. The proposed method first constructs the propagation tree. It then captures the propagation behaviours features which fall into three types: minimum value, standard deviation and average value. Finally, four models were applied: SVM, logistic regression, Adaboost and random forest. 

Zhou et al. \cite{Makinouchi2018} explored time-series dynamic characteristic of sockpuppet network. The authors observed that, as soon as the account is blocked, a puppetmaster tries to restore its prior social relationships quickly to maintain its impact. A sockpuppet detection technique was proposed based on this investigation. First, a weight representation technique was designed to record the dynamic growth of the social relationships between the two users \(u_i\) and \(u_j\). The weight \(w_{u_i}^{u_j}\) was calculated, as shown in Equation \ref{zhooo122}, where \(\lambda\) is the proportion of the time series in the weight and \(K(.)\) is the time series of two nodes forming edges in the social network. Next, the detection technique of sockpuppets is formalized as a similarity time series analysis problem. Shortest path hops and the weighted sum of edges were applied to show the degree of the relationships among  accounts and the degree function \(\phi (u_i, u_j) = \frac{w_{u_i}^{u_j}}{|p_{u_j}^{u_i}|^2}\), where \(p_{u_i}^{u_j}\) is a shortest path from node \(u_j\) and \(u_i\). The interaction graph similarity was calculated, as shown in Equation \ref{zhooo2}, where \(V\) is a set of nodes and \(x_i\) is the \(i\)-th element of \(X\). 
\begin{equation}\label{zhooo122} w_{u_i}^{u_j} = (1 - \lambda) * F(u_i,u_j) + \lambda * K(u_i,u_j) \end{equation} 
\begin{equation}\label{zhooo2} H(u, v) = \frac{\sum_{i=1}^{|X|}(\phi (u, x_i) + \phi (v, x_i) )}{|V_u \cup V_v|} \end{equation} 

Wang et al. \cite{Wang2018} investigated three different approaches to predict if multiple accounts belonged  to the same user. The authors investigate Katz similarity \cite{Katz1953} based unsupervised learning, Katz similarity-based supervised learning and graph embedding based semi-supervised learning. First, the authors constructed a bipartite graph\( G\). The similarities between different accounts is calculated based on\( (I - \beta \ M)^{-1} - I\)  where\( M\) is the representation of adjacency matrix of\( G\), \( \beta\) is a scalar smaller than\( 1/||M|| \ 2\) to ensure convergence and\( I\) is the identity matrix. Next, the similarity matrix is computed to measure account closeness in the graph. The account belongs to the same user if the similarity is bigger than a threshold percentile. The Katz similarity-based semi-supervised learning yields poor prediction accuracy. Therefore, graph embedding was used to feed extra data into the model because it can extract more data from the graphs than Katz similarity and can thus achieve a higher prediction accuracy.  The clustering-based approach was utilized to improve scalability. The shortfall of this technique is that its performance
relies on the selection of an appropriate threshold. This is impractical and inefficient when dealing with diverse activities.

Nazir et al. \cite{nazir2010ghostbusting} focused on detection and classification in creating multiple phantom gaming profiles to obtain a strategic power in social games. An approach was proposed to detect and potentially eliminate phantom user profiles. Empirical characterization was performed to identify distinctive traits that could help to distinguish between phantom and real profiles. Conditional cumulative distributions associated with genuine and phantom profiles are computed and compared. An SVM was then applied to determine the authenticity of profiles by generalizing the information of genuine and phantom user profiles. An SVM was implemented for classification.

A downside of the similar-orientation network-based techniques is that they only focus on detecting the sockpuppet accounts which have similar behaviours. An adversary can mimic genuine accounts' behaviours, which may lead to the failure of these techniques.

Table 4 in Appendix presents an overview and comparison of the aforementioned papers based on the goals, techniques, features and datasets.

\subsubsection{Social Botnet Detection Techniques}
Many researchers have proposed advanced approaches to automatically detect social botnets or to differentiate between bots and humans. So far social media providers have not employed adequate strategies to minimize social botnet phenomena \cite{Ferrara2016}. 
We categorize these techniques into three classes: \textit{text mining-based techniques, machine learning-based techniques and deep learning-based techniques.} 
\paragraph{Text Mining-Based Techniques}These techniques base on text categorization, text clustering and concept/entity extraction. These techniques are also capable of processing a large amount of unstructured data. 

Dickerson et al. \cite{Dickerson2014} proposed a sentiment-aware architecture to detect Twitter bots account, called SentiBot, where the novel semantic features of the sentiment of a tweet in bot and human users were analyzed. SentiBot first extracts the Twitter sentiment of users' tweets. It identifies a set \(\ U_0\) of users based on the topic of interest within a selected time. It then calculates the score of the sentiment  \(\ SS(d,u,t)\) of user \(\ u\) on topic \(\ t\), across all tweets on topic \(\ t\) posted by the user on that day \(\ d\). It examines users' profiles to construct a Twitter follower network subgraph produced by the set of \(\ U\) users. It identifies the set \(\ U_1 = \) \{\(\ u' |(\exists u \in U_0) \) such that either \(\ u\) follows \(\ u'\) or \(\ u'\) follows \(\ u\)\} for each user in \(\ U_0\). It also identifies the set \(\ U_2 = \) \{\(\ u'' |(\exists u' \in U_1) \) such that either \(\ u'\) follows \(\ u''\) or \(\ u''\) follows \(\ u\)\}. The complete users' set is defined as \(\ U= U_0 \cup U_1 \cup U_2\).  Next, SentiBot extracts a user profile by looking at that user’s position in the network, the user’s tweet and the user’s behaviour. SentiBot constructs a database. The rows of the constructed database correspond to users. The columns of the constructed database correspond to contextual variables' set.  Six classification techniques were utilized: SVM, Adaboost, Gaussian naive Bayes, random forests, gradient boosting and extremely randomized trees. 

Zhang et al. \cite{Zhang2018b} investigated the advantages and the efficacy of spam distribution and digital-influence manipulation based on the social botnet on Twitter. The corresponding countermeasures were proposed and effectively evaluated. First, the authors proposed a defensive technique against spam distribution. In this attack, retweeting trees are exploited to spread spams by the adversary. For example, the adversary posts tweets contained malicious URLs. The user who has a participating history in spam distribution is tracked.  If the user's behaviour exceeds a certain threshold, the user will be suspended. Each user \(\ v\) has a spam score \(\ s_v\). When the user \(\ v\) retweets spam, the spam score \(\ s_v\) is updated. Once user \(\ v\) exceeds the predefined threshold of the \(\ s_v\), the user \(\ v\) is deemed a spammer and suspended. The spam score is defined as \(\ s_v=s_v + \gamma^d\), where \(\ d\) is the number of retweeting hops between the spam source to \(\ v\) and \(\ \gamma \leq 1 \) is the attenuation factor of distance. Second, the authors proposed a defensive technique against digital-influence manipulation. The number of actions is considered by vendors of digital-influence software. The digital influence of a user is evaluated by the audience size. Therefore, the digital influence of a user is exposed to social bots. A new digital-influence measurement scheme is proposed.  This scheme's  concept is to search for sufficient honest users. Digital-influence scores for other users are calculated based on honest users' actions.  Given a social network comprising a set of user  \(\ V\) and all their actions (such as retweeting, following, replying  and mentioning). The scheme first searches for honest users which are a subset \(\ V^* \subseteq V\) of users. The digital-influence score of each user \(\ u\) is then defined. 

Text mining-based techniques require extensive textual data. Social bots nowadays can emulate genuine humans' content production behaviours, including temporal spikes of content production and circadian exemplars of daily activities. Social bots also can automatically interact with humans by answering their questions and commenting on their posts \cite{hwang2012socialbots}. These new features greatly challenge the text mining-based techniques.

\paragraph{Machine Learning-Based Techniques}
These techniques mostly base on feature extraction. Researchers encode and adopt behavioural patterns to differentiate between human and bots \cite{Ferrara2016}.  

Khaled et al. \cite{Khaled2019} proposed a classification algorithm, SVM-NN, to detect fake Twitter accounts and bots. Neural network (NN) was employed with decision values arising from an SVM. The proposed method consists of four stages: data pre-processing, feature reduction, data classification and accuracy comparison. First, the provided data are transformed into a proper form for classification. Four feature reduction techniques were later adopted: Principal Component Analysis, Wrapper Feature Selection using SVM, SpearmansRank-Order Correlation and Multiple Linear Regression. The authors then proposed a new algorithm where the NN model is trained by the SVM trained model decision values, and the NN model is tested by the SVM testing decision values. 

Balestrucci et al. \cite{Balestrucci2019} investigated the relationship between bots and legitimate human accounts. The authors sought to detect credulous accounts on Twitter starting with those involved in any social relationship with a bot, and proposed an approach to automatically rank them based on their gullibility. The proposed approach first crawls Twitter data to retrieve user information. The user data are then transformed into a suitable format for querying the inspector. Next, the proposed approach performs human-bot classification. The experimented candidate models include decision trees, random forest, RIPPER and neural networks based on multiple feature subsets, i.e., user, friends, network, temporal, content and sentiment-based features. 

Chu et al. \cite{Chu2012} investigated human, cyborg and bot accounts' classification on Twitter. First, a series of measurements to differentiate between human, cyborg and bot accounts were conducted. These accounts are differentiated in tweeting behaviours, tweet content and account properties. 
An analysis of the collected data was performed to discover a collection of features.  The authors proposed a classification method based on the measurements' results. This method has four components: entropy, spam detection, account properties and decision-maker. The entropy-based component detects the periodic and regular timing which is an automation indicator. The entropy rate measures a process complexity. The authors used the corrected conditional entropy to calculate entropy rate. The corrected conditional entropy (\(CCE\)) is calculated as shown in Equation \ref{eq:ccc}, where \(\ perc(X_m)\) is the percentage of unique sequences of length \(m\), \(EN\) is the entropy with \(m\) fixed at 1 and \(CE\) is the conditional entropy. Moreover, the spam detection component checks the text patterns for spam content. The probability of a message \(\ M\) being spam is determined by Baye's theorem, as shown in Equation \ref{eq:ccc1}. The account properties component uses account properties for deviations detection from normal. Finally, Random Forest was applied in the decision-maker part. It takes features combination as an input to classify anonymous accounts as human, cyborg, or bot. 
\begin{equation}
\label{eq:ccc} 
\begin{aligned}
CCE(X_m|X_1,...X_{m-1}) = CE(X_m|X_1,...X_{m-1})+perc(X_m) \cdot EN(X_1) 
\end{aligned}
\end{equation} 
\begin{equation}
\label{eq:ccc1}
\begin{aligned}
P(spam|M) = \frac{P(M|spam)P(spam)}{P(M)} = \frac{P(M|spam)P(spam)}{P(M|spam)P(bot)+P(M|not spam)P(not spam)}
\end{aligned}
\end{equation} 

The performance of machine learning-based techniques heavily relies on the quality of feature engineering, where feature engineering is one of the most challenging and time-consuming tasks. There is also no attempt to explore social botnets in a new context such as interference of bots with public discourse, etc.

\paragraph{Deep Learning-Based Techniques}
Kudugunta and Ferrara \cite{kudugunta2018deep} proposed a contextual long short-term memory (LSTM) approach for detecting bots based on their tweets. The user metadata is extracted and fed as input to LSTM. This approach has two classification phases: account-level and tweet-level bot detection. In the account-level bot detection phase, the authors used a synthetic minority oversampling technique (SMOTE) because it is a dominant approach which has been achieved in many fields. SMOTE with two data enhancements are combined. The two data enhancements are Tomek Links and Edited Nearest Neighbors (ENN). SMOTE was applied to the whole dataset. Tomek Links and ENN are only applied to the majority class. The tweet-level bot detection phase determines if the tweet's author is a bot. The LSTM model was used to overcome the limitations of traditional techniques of Natural Language Processing tasks. A pre-trained set of Global Vectors for distributed word representation (GloVE) was employed. GloVE converts the text of the tweet into an amenable formation for processing via LSTMs. GloVE is a log-bilinear regression model which uses local context windows and global matrix factorization to efficiently learn the natural language substructure, by training on word co-occurrences. Similar to the limitation of the machine learning-based techniques, this technique detects social botnets based on only the account and tweet levels. Exploring social botnets from other contexts has not been conducted.

Table 5 in Appendix presents an overview and comparison of the aforementioned papers based on the goals, techniques, features and datasets. 
\subsubsection{Other Fake Profile Detection Techniques}
This section focuses on detection techniques for all minor categories of fake profile that are not as dominant as the other three main categories of fake profile. An example of the minority of fake profiles is the creation of an account by providing fake user gender which is considered a profile attribute deception. We categorized these detection techniques based on their targets.
\paragraph{General Fake Account Detection.} These techniques target unclassifiable fake accounts. An example is the analysis of the characteristics of fake accounts. We also consider the fake accounts who publish or spread spams. Spams are malicious and deceptive content and links embedded in advertisements, user-submitted comments and posts.

Walt and Eloff \cite{VanDerWalt2018} investigated the applicability of existing bot detection features to detect fake accounts created by human. The proposed detection approach comprises seven steps. The data were gathered and cleaned from bot accounts. The authors manually created 15000 fictitious accounts. Chi Square and Whitney-U tests were then used to validate the data. Fictitious accounts were later injected into the original dataset. Engineered features collected from previous research were employed. Three machine learning models were applied: random forest, SVM and Adaboost for fake account detection. This technique heavily relies on the quality of feature engineering.

Caruccio et al. \cite{Caruccio2019} proposed a novel approach to differentiate real users from fake users. The proposed technique extracts Relaxed Functional Dependencies (RFDs) from the social network dataset. The proposed technique classifies Twitter users into real, verified and fake ones. The proposed technique is able to effectively distinguish fake users from real users. It also helps to find significant patterns distinguishing human behaviours.
This technique is currently only applied in the Twitter dataset.

Gurajala et al. \cite{Gurajala2015} analyzed 62 million Twitter user profiles to examine the characteristics of fake account creation. A map-reducing approach and a pattern recognition approach were employed for fake accounts detection. A highly reliable fake user subset is detected. The users were grouped based on an update-time distribution filter, patterns in their screen names and matched attributes of multiple profiles. The authors observed that profile registration time and the fake profile URLs set have behaviours which are distinct from the ground truth dataset. The limitation of this technique is that it can only detect a relatively small proportion of fake users in the experiment.

Kumari and Rathore \cite{Kumari2018} presented a model to detect potential fake accounts based on profile information and account activities. The proposed model was implemented via a Facebook application called SocialMedia. The proposed model relies on five features: number of mutual friends, family, age group, average likes and average comments. The proposed model then calculates the ``Trust Weight'' among a user\(\ u\) and his/her friends\(\ v\) based on Equation \ref{kumari}. The user logs into the SocialMedia application and permits the SocialMedia application to access their information. The SocialMedia application extracts the required features. The trust among the user and his/her friends is calculated. Finally, the SocialMedia application sorts the friends based on the Trust Weight. 
\begin{equation}
\begin{split}
\label{kumari}
&Trust \ Weight \ (u,v) = No. \ of \ mutual \ friends (u,v) + Average \ likes +  Average \ comments \ \\ 
&+Age \  group + 100 \times (Is Family (u,v))
\end{split}
\end{equation}

Wang \cite{wang2010don} proposed a prototype system for suspicious accounts detection and identification on Twitter. A directed social graph model was proposed in order to examine (friends and follower) relationships among users. The main objective is to employ machine learning approaches to automatically differentiate legitimate accounts from spammers. In the social graph model, each node represents a user account. If a specific user follows another user, then they will be represented by a directed edge. Four types of relationships are defined: 1) followers which represent the people who are following you, 2) friends who are the users to whose updates you are subscribed, 3) strangers, if there is no connection among users and 4) mutual friends if two users are friends with each other, or who are following each other. Content and graph-based features are extracted based on Twitter's unique characteristics. SVM, decision tree, naive Bayes classifiers and NN were experimented. The author considered each account as a vector \(\ X\) with feature values. The objective is to allocate each user to one of two groups \(\ Y\): non-spam and spam. Each conditional probability is estimated independently. The posterior probability is calculated for each class to classify Twitter accounts using Equation \ref{eq:4}. 
\begin{equation}\label{eq:4} P(Y|X) = \frac{P(Y)\prod_{i=1}^{d} P(X_i|Y)}{P(X)} \end{equation}

Zhu et al. \cite{Zhu2012} proposed a Supervised Matrix Factorization method with Social Regularization (SMFSR) to detect spammers. They proposed a novel framework which extracts social activities features to predict spammers. A matrix factorization model was used to factorize the activity matrix into the latent matrix of users\(\ U\)  and the latent matrix activities\(\ V\). The learning process of the latent feature is conducted by the label information and its relationship with the social graph. The input is the produced features for a spammer classification model. SMFSR is defined based on Equation \ref{eq:zzzz}, where\(\ \alpha\) is the tradeoff coefficient between the classification loss and factorization loss,\(\ \bf w\) is the coefficient vector for user latent factors,\(\ R_s\) is a social regularization which helps to avoid the overfitting,\(\ h\) is the smoothed hinge loss,\(\ \ell\) is the number of labelled data and\(\ {\bf w}^T U_i\) is a linear classifier which can be applied to the latent user features.  
\begin{equation}\label{eq:zzzz} 
\begin{aligned}
\jmath_s(U,V,{\bf w} ) =  \sum_{aij \in I_{ij}} (a_{ij}- \sum_{f=1}^K U_{if}V_{if})^2  + \frac{\alpha}{2} \sum_{i=1}^\ell h (y_i({\bf w}^T U_i))+\frac{\lambda_w}{2}||{\bf w}|| _2^2 + \frac{\lambda_s}{2}  R_s + 
\\ \frac{\lambda_f}{2}(||U|| _F^2 +(||V|| _F^2 + ||{\bf W}|| _F^2) 
 \end{aligned}
 \end{equation}

Singh et al. \cite{Singh2016} analysed the behaviours of Twitter accounts which tweet pornographic content. The main goal of the work is to distinguish between legitimate accounts and spammers based on confirmed behavioural characteristics driven by Twitter’s policies. NodeXL was employed to gather and analyse the data from Twitter. NodeXL is an application which explores the data of a social network. The collected data comprises pornography-related tweets based on six keywords: nude, pornography, adult sex, boobs, lesbian and anal/ass. The authors classified the pornographic users into genuine or pornographic spammers under the spammers' category based on behavioural analysis.  five classification algorithms were experimented: random forest, Bayes network, logistic regression, Adaboost and J48 Classifier. 

Benevenuto et al. \cite{Magno2014} proposed a four-step approach to detect spammers on Twitter. They first collected approximately 54 million accounts, 1.9 billion links and almost 1.8 billion tweets. Secondly, they manually categorized the collected accounts into non-spammers and spammers. The collected data focused on the most trending events in 2009.
The characteristics of tweet content and user behaviours were studied to distinguish spammers and non-spammers. Finally, an SVM was implemented to identify spammers. 

A fundamental limitation of these techniques is that they ignore the fact that attackers can mimic a genuine profile. This causes the incapability of these techniques in differentiating between genuine and fake profiles.

\paragraph{Profile Attribute Deception Detection.} These techniques are targeting accounts with fake profile attributes.
Alowibdi et al. \cite{Alowibdi2014a} proposed a technique for detecting user gender deception in Twitter. This technique compares indicators of gender extracted from various profile attributes including first name, username and layout colours. Each indicator strength and accuracy of all possible values were examined for each indicator through considerable experimentations. Trends of male and female users are defined, both based on the overall accuracy of each profile attribute and the relative strength of the value of each attribute for a specified user. The Bayesian classifier was applied to the weighted average of attributes for each user. The factor of male trending $m$ of each user profile is calculated using Equation \ref{eq:alowi}, where  $w_f$,  $w_u$ and   $w_c$ represent the relative weights of the three indicators of gender and  $s_f$,  $s_u$ and   $s_c$ represent the sensitivity of a user’s feature for a given indicator. The authors then divided the user profiles into five groups based on the calculated male index  $m$ after calculating the index of male trending  for each user profile in the dataset. The user profiles are defined based on $m$ values. Firstly, strongly trending female and weakly trending female profiles fall in the range \(\ 0 \leq m \leq \mu -2 \sigma \), and \(\ \mu -2 \sigma < m \leq \mu - \sigma\) respectively. Secondly, strongly trending male and weakly trending male are considered in the range \(\ \mu +2 \sigma \leq m \leq 1\) and \(\ \mu +2 \sigma \leq m < \mu + 2 \sigma \) respectively. 
This technique ignores the fact that social media users can fabricate user names and profile colours to misrepresent their gender. 
In addition, the research has not explored how cultural differences affect the prediction. For example, a certain profile colour might be associated with different genders in different cultures.
\begin{equation}
\label{eq:alowi} m = \frac{w_f \ . s_f \ + w_u \ . s_u \ + w_c \ . s_c}{w_f \ + w_u \ + w_c}
\end{equation} 

\paragraph{Fake Account Detection in Sub-Communities.} The following technique targets fake accounts in sub-communities.
Tsikerdekis et al. \cite{Tsikerdekis2017a} proposed a proactive approach to detect identity deception for sub-communities. Sub-communities exist in larger communities, for example, Subreddits and Facebook groups. The proposed approach utilizes social network data and specifically a common contribution network. The social network formation bases on common contributions made by members. The edge formation between two users is made because they participated in the steam Subreddit. The structure of the network indicates shared mutual interests among users. An algorithm was utilized to create a view of a sub-community's network, which does not include isolated nodes. The authors employed  a supervised machine learning method. The computational efficiency of this technique in large social networks has not been explored. 

Table 6 in Appendix presents an overview and comparison of the aforementioned papers based on the goals, techniques, features and datasets.

\subsection{Identity Theft Detection Techniques}
\subsubsection{Compromised Account Detection Techniques}
We classified the compromised account detection techniques into \textit{behaviour-based, text-based and machine learning-based techniques. }
\paragraph{Behaviour-Based Techniques}
These techniques significantly depend on the sufficiency of behaviour records, such as keystroke, user-generated content and clickstream.  It also depends on collecting accounts' suspicious behaviour patterns to differentiate between normal and suspicious accounts. The compromised account does not behave in the manner of a normal account. This can be detected by capturing the account's behavioural patterns.  

Wang et al. \cite{Wang2018a} proposed a Composite Behavioral Model. It is a joint probabilistic generative model based on Bayes networks. The proposed model bases on two different behaviour spaces: check-in location in offline behaviour space and user-generated content in online behaviour space.  Each composite behaviour is denoted by a triple-tuple\(\ (u, v, D)\). The chance of user\(\ u\) visiting venue\(\ v\) and tweeting a tip online with a set of words\(\ D\) is computed.  A relative anomalous score\(\ S_r\) was devised by using Equation \ref{wangggg} to measure the occurrence rate of each composite behaviour\(\ (u, v, D)\) of different levels of activity of different users. 40 users were randomly selected to estimate the relative anomalous score for each composite behaviour. The limitation of this approach is that it can only detect identity theft after the access control of the account is broken.
\begin{equation}\label{wangggg} S_r(u, v, D) = 1 - P(u|v,D) = 1 - \frac{P(v,D|u)P(u)}{\sum_{u^\prime}P(v,D|u^\prime)P(u^\prime)} \end{equation}

\paragraph{Text-Based Techniques}
These techniques convert an unstructured text into directly usable data. It includes entity relation modelling, entity extraction, text categorization and sentiment analysis. It also helps to process a large amount of unstructured data.  

Kaur et al. \cite{kaur2018authcom} performed authorship verification to examine the posted tweets' authorship. They analysed various textual features, such as stylometric, n-grams, folksonomy and Bag of Words (BOW) features. Different feature selection techniques were examined, including Chi-squared, correlation feature selection, AHP-TOPSIS, Fisher score, Gini Index and t-score. AHP-TOPSIS compares the relative closeness of each feature while the other feature selection techniques only consider feature-feature or feature-class correlation.  Statistical analysis and classification techniques were applied to calculate various performance parameters. 

Seyler et al. \cite{Seyler2018} proposed a framework for detecting compromised accounts by utilizing statistical text analysis features. The framework bases on the observation that a user’s textual writing is measurably different from a hacker's textual writing. The tweet space of a user is divided into two non-overlapping sets\(\ M^{spam}\)  and  \(\ M^{user}\) to capture the discrepancy between language usage. The difference between\(\ M^{spam}\) and\(\ M^{user}\) is measured using Kullback-Leibler-divergence\(\ D_{KL}(P,Q) = \sum_{i}P(i) log (\frac{P(i)}{Q(i)})\). KL-divergence is a  variation measure within two probability distributions\(\ P\) and\(\ Q\) \cite{kullback1951information}\cite{zhai2016text}. The maximum, minimum, mean and variance of the sampled KL-divergence scores are considered as the features of the classifier. 

A common pitfall of the text-based techniques is that they cannot detect an adversary who can mimic a victim's posting behaviour.

\paragraph{Machine Learning-Based Techniques}
We present various machine learning-based social media compromised account detection techniques. Machine learning-based techniques generally base on feature-based methods which capture abnormal behaviours \cite{Ferrara2016}.  

Singh et al. \cite{Singh2018} proposed a novel real-time streaming system to detect spammers. The proposed system bases on various features: trust, user and content. Unlike previous works that brought up the spam and compromised account detection, the authors developed an integrated approach able to detect various types of malicious users. The proposed system follows these steps: data collection, robust features identification and classifier selection. Two types of data were used. First, collected labelled data were used to train the classifier. Second, crawled random data were used to test the classifier. Three categories of features were adopted in the identification of robust features step, including user-based features like account ages, trust-based features like trust scores, and content-based features like links in the tweets. Five machine learning classification techniques were utilized in the final step, namely Bayes network, logistic regression, random forest,  Adaboost and J48. 

Cao et al. \cite{Cao2014} proposed a detection system called SynchroTrap for suspicious accounts. It detects both fake and compromised accounts. It can detect a sizable group of malicious users who act in loose synchronicity. Account actions are compared within a certain time. Users are then categorized based on the actions' similarities. SynchroTrap uses adversary’s network resources constraint. For example, IP addresses number, or the adversary’s target, to reduce the combination comparison. SynchroTrap was developed as an incremental processing technique on Giraph and Hadoop. 

Villar-Rodríguez et al. \cite{Adhianto2010} proposed a novel approach for identity theft detection based on connection time traces. The  approach strategy aims to synthetically construct traces of connection time. The constructed traces indicate identity theft events over the user account. User connection time is recorded. The behaviour of the user is modelled as the duration of user connection time and when the connection is started. These patterns are then fed to an SVM classifier.

Similar to other machine learning-based techniques, machine learning-based comprised account detection techniques relies on the quality of feature extraction. The bots' interference with public disclosure has not been considered in the existing detection techniques.


Table 7 in Appendix presents an overview and comparison of the aforementioned papers based on the goals, techniques, features and datasets.

\subsubsection{Other  Identity Theft Detection Techniques}
This section provides an overview of the other types of identity theft detection techniques. These techniques focus on those scenarios that do not target compromised accounts. Such an example includes creating an account using  an individual's stolen personal information on a social media platform, where this individual does not have an account on this social media platform. 

Conti et al. \cite{Conti2012} analysed social network graphs to detect an impersonated real person who had no online profile. This technique aims to examine the social network temporal evolution and real account profile characterization. The authors used the interaction of a social network feature and its graph properties. The features consist of overtime evolution of the friends' numbers of the social network, the interaction of a real social network, and overtime evolution of the social network graph structure. A ``sensing'' application on Facebook was proposed to collect the required statistical information from a user profile. The application generates dynamic information. Each stream is linked to a timestamp when it has been reported in the wall of the user. The soundness of this technique is challenged by the small number of user-profiles (i.e., 80) studied in the evaluation.

He et al. \cite{He2014} proposed an identity protection scheme. The authors designed a scheme whereby users’ personal information is kept public without identity being stolen. The identity theft protection scheme consists of three approaches: Challenge, Login Account as An Identifier and Friend Network Similarity. In Challenge, users send messages to each other to confirm the new relationship in the new social media. In Login Account as an Identifier, users check whether the new friend is available in the other social media platforms. Friend network similarity is calculated by using a similarity measure. Equation \ref{eq:hh} defines the friend network similarity between two accounts (\(\ v_{t}^{n}\), \(\ v_{c}^{m}\)) in a social network. Equation \ref{eq:hh1} defines the probability of \(\ v_{c}^{m}\)'s validity as P. \begin{equation}\label{eq:hh} S(v_{t}^{n}, v_{c}^{m}) = \frac{|CF (v_{t}^{n}, v_{c}^{m})|}{|EX(G_m, v_{t}^{n}.fl)|} \end{equation} 
\begin{equation}\label{eq:hh1} P(S(v_{t}^{n}, v_{c}^{m}))= \lambda \end{equation}
where \(\ |CF (v_{t}^{n}, v_{c}^{m})|\) represents the number of common friends between \(\ v_{t}^{n}\) and \(\ v_{c}^{m}\). \(\ |EX(G_m, v_{t}^{n}.fl)|\) represents the number of a friend in social network\(\ G_m\). Next, a value \(\ \alpha\) is obtained after calculating the friend similarity. If the \(\ \alpha\) increases, the similarity between \(\ v_{t}^{n}\) and \(\ v_{c}^{m}\) also increases. \(\ \alpha\) is evaluated to get the probability of \(\ v_{c}^{m}\) being real in a social network. 

The following work targets analyzing the behavioural patterns of the attackers. Zaeem et al. \cite{NokhbehZaeem2017} analysed and described thieves' and fraudsters' behavioural patterns and resources.
They proposed an automated solution which employs text mining to extract important information from identity theft stories and articles.  A pipelined system takes the news of identity theft as input. It outputs the analytics to help understand the process of identity theft. Search engines were used to collect articles from annual identity theft reports. The authors pre-processed the news stories by eliminating unused language. Part-Of-Speech and Named Entity Recognizer tagger functions were utilized to extract the named entities. Next, each story represents an identity theft record, which is used to analyse various aspects of identity theft. The analysis result shows an increase in the understanding of identity threat behaviours. No corresponding detection technique has been proposed.

Table 8 in Appendix presents an overview and comparison of the aforementioned papers based on the goals, techniques, features and datasets.

\subsection{Identity Cloning Detection Techniques}
We classified the detection approaches of identity cloning accounts into two main classes: \textit{user profile-based techniques and graph-based techniques.}
\paragraph{User Profile-Based Techniques}
These techniques rely on user profile information and collecting accounts' suspicious behaviour patterns to assess account cloning.

Kontaxis et al. \cite{Kontaxis2011} proposed a methodology that can be employed by account holders to see if they have fallen victim to identity cloning. The methodology bases on identifying any user profile information that can uniquely identify the real person's identity. It has three phases: information distiller, profile hunter and profile verifier. The information distiller is responsible for extracting user-identifying information from the real user's profile and analysing that profile to find which information is considered rare or user-specific to distinguish the user. The profile hunter is responsible for processing the user information across online social networks to find if there are similar user profiles. The profile verifier is responsible for processing user information and extracting the information in the harvested social profiles. It then calculates the similarity scores between those profiles and the real one. 

Devmane and Rana \cite{Devmane2014} designed a methodology to detect identity cloning attacks in the same site and cross-site. The proposed methodology first extracts user profile information and then searches for the account profile on the same and cross-site social media. Next, a similarity index is calculated to detect the cloned profiles. 

Jin et al. \cite{Jin2011} analysed and characterized the behaviours of the identity clone attacks. Firstly, the authors computed the attribute similarity for two identities (\(\ S_{att}\)). Given a candidate public profile\(\ P_C\) and a victim public profile\(\ P_v\), they calculated the total number of similar attributes (\(\ SA_{CV}\)). The authors then defined the attribute similarity as shown in Equation \ref{jjj2}, where\(\ A_c\) and\(\ A_v\) represent the number of attributes of\(\ P_C\) and\(\ P_v\), respectively. Two profile similarity metrics were proposed to detect suspected profiles: Basic Profile Similarity (BPS) and Multiple-Faked Identities Profile Similarity (MFIPS). It is assumed that the adversary does not forge the victim's friends in the BPS. The BPS is shown in Equation \ref{jjjj}, where\(\ S_{bfn}\) is the friend network similarity of\(\ P_C\) and\(\ P_v\) and\(\ \kappa\) and\(\ \chi\) are the parameters to balance the effect of attribute similarity and friend network similarity. The MFIPS is formalised based on the assumption that the adversary forges the victim's friends. The MFIPS is calculated as shown in Equation \ref{jjjj1}, where\(\ S_{mfn}\) is the multiple-faked identities profile similarity. 
The disadvantage of this technique is that the proposed framework cannot detect the accounts cloned from other social media platforms.
\begin{equation}\label{jjj2} S_{att} (P_c,P_v) = \frac{SA_{CV}}{\sqrt{\mid A_c\mid \times \mid A_v \mid}} \end{equation}
\begin{equation}\label{jjjj} S_{BPS} (P_c,P_v) = \frac{\sqrt{(\kappa S_{att})^2+(\chi S_{bfn})^2}}{\sqrt{\kappa^2 + \chi^2}} \end{equation}
\begin{equation}\label{jjjj1} S_{MFPS} (P_c,P_v) = \frac{\sqrt{(\kappa S_{att})^2+(\chi S_{mfn})^2}}{\sqrt{\kappa^2 + \chi^2}} \end{equation}

A common shortfall of these techniques is that their performance in a real-world environment has not been fully explored.


\paragraph{Graph-Based Techniques}
Kamhoua et al. \cite{Kamhoua2017}  overcame identity cloning attacks by matching user profiles across multiple social media. A three-step method was proposed based on supervised machine learning approaches. The used features are extracted from both friend request information and friend lists of the users. The authors used a binary classifier for ranking the probability that friend request of multiple users from various users can be colluders. The proposed method first collects profile information of the users' friend lists request. It then searches for a similar profile based on the collected information across the same or other social media. The proposed method verifies each identity of the user-friend request. Finally, the proposed method returns whether or not the friend requests are colluders. A hybrid-based string-matching similarity algorithm was used to find profile similarity in this work. FuzzySim is a modified similarity metric, as shown in Equation \ref{kamboo}, where\(\ A\) and\(\ B\) are two strings,\(\ \bigcap_{avrg \ \theta} \) was used as an average weighted threshold and\(\ max(|A|,|B|)\) is the max string length. It overcomes the problem of asymmetry of the Monge-Elkan \cite{monge1996field}. A downside of this approach is that it has not been evaluated on a real-world dataset. 

\begin{equation}
\begin{split}
\label{kamboo}
&FuzzySim_{MongeElkan}(A,B) = \frac{1}{max(|A|,|B|)} \sum_{i=1}^{|A|} \bigcap_{avrg \ \theta} \{sim' (a_i,b_j)\}_{j=1}^{|B|}
\end{split}
\end{equation}

Table 9 in Appendix presents an overview and comparison of the aforementioned papers based on the goals, techniques, features and datasets.

\section{Discussion}
The above literature on detection techniques of social media identity deception is analysed in five dimensions: domains, detection techniques, features, evaluation metrics and datasets.
\subsection{Domains}
Detection techniques of identity deception on social media platforms have focused on various domains. Identity deception detection domains are divided into generic and singular domains. The generic domain refers to all types of social media while the singular domain targets specific social media platforms. Approximately 60\% of the literature focuses on detecting identity deception from generic social media, as shown in Table 10 in Appendix. Twitter is the most dominant singular domain, used by a third of the studies. The literature that targets Twitter mostly focuses on detecting fake profiles (social botnets and others). Online discussion forums and Wikipedia are the general targets of sockpuppet detection techniques. Facebook is the focus of 8\% of the surveyed studies. The existing research also focuses on detecting identity deception in Instagram and news websites. 

It can be inferred from Table 10 in Appendix that 80\% of the literature concerning Sybil detection techniques has focused on the generic social media domain, while approximately 60\% of the research on sockpuppet detection techniques has focused on the same scope. The most targeted singular domain for sockpuppet detection techniques is the online discussion forums. 83\% of the social botnet detection literature focuses on Twitter. 66\% of the literature for compromised account detection techniques is focused on generic social media. Twitter is the most focused singular domain for compromised account detection techniques. Identity cloning detection techniques are focused only on the generic social media domain.
\subsection{Detection Techniques}
Graph-based techniques and machine learning-based techniques are the most dominant techniques for identity deception in social media, as shown in Table 11 in Appendix.  26\% of the literature adopts machine learning-based techniques. Verbal behaviour analysis, non-verbal behaviour analysis and similar-orientation networks are only leveraged in sockpuppet detection techniques. 43\% of the literature on sockpuppet detection techniques adopts a similar-orientation network. Additionally, the key management-based technique is only considered by one study on Sybil detection techniques. Surprisingly, none of the literature to date has employed deep learning-based techniques except for social botnet detection techniques. 

As shown in Table 11 in Appendix, 50\% of the literature concerning Sybil detection techniques has adopted graph-based techniques, followed by machine learning-based techniques. 40\% of the literature on social botnet detection techniques deploys techniques based on machine learning and text mining, respectively. 50\% of the literature applies textual-based techniques for compromised account detection. Identity cloning detection techniques are divided into graph-based techniques (1 study) and user-profile techniques (3 studies).
\subsection{Features}
Identity deception detection techniques features are categorized into eight classes in terms of their nature, namely, content, user profile, time, graph, distribution structure, community and propagation tree, as shown in Tables 12 and 13 in Appendix. 33\% of the literature relies on content-based features.  Those features depend on content of user posts, such as the number of users' posts and writing styles of the posts. The most dominant feature is the number of posts, which has been used by 61\% of the studies that rely on the content-based features. The user profile features incorporate information of users' profiles, such as the number of a user's friends and a user's reputation. The user profile features have been adopted in 24\% of the studies. 20\% of the literature that relies on the user profile-based features has employed the number of followers. Time-based features are adopted by 18\% of the studies. The time-based features focus on a user's time on social media, such as connection time and the time between posts. The time between posts and account age are the most significant features used in the time-based features. Graph-based features are used by 9\% of the studies. Additionally, two studies only use features based on structure, community and distribution. The structure features include the total number of revisions and whether a user uploads a file. The community features base on the user status on social media, such as whether the user has been reported. The distribution features base on the patterns of the user-entered data, such as name, education and company. Finally, the propagation tree-based features are only used in one study.

As shown in Tables 12 and 13 in Appendix, the literature on Sybil attack detection techniques typically focuses on the following features: graph-based features (3 studies), content-based features (2 studies), user profile-based features (2 studies) and time-based features (1 study). sockpuppet detection techniques apply all the types of features identified in Tables 12 and 13 in Appendix. The content-based features are the most used features in sockpuppet detection techniques. 50\% of the literature on social botnet detection techniques uses content-based features. The number of user mentions is the most dominant features used in social botnet detection techniques. Furthermore, compromised account detection techniques employ content-based features in 40\% of the studies.  The user profile-based features, time-based features and distribution-based features (1 study each) are also used by compromised account detection techniques. Surprisingly, identity cloning detection techniques only rely on user profile-based features.
\subsection{Evaluation Metrics}
A variety of evaluation metrics are employed to investigate the validity of identity deception detection techniques in social media, as shown in Table 14 in Appendix. Accuracy and Precision are the most dominant evaluation metrics, which are appeared in 50\% of the studies. The second most concerned evaluation metric is Recall, which is used by 41\% of the studies. 32\% of the literature uses F-measure. False negative rates and false positive rates are adopted by 35\%  and 16\% of the studies, respectively. The less frequently utilized evaluation metrics include Matthews correlation (3 studies), receiver operating characteristic (ROC) (2 studies), coefficient of variance (CV) (2 studies), false rejection rate and false acceptance rate (FAR) (1 study each).

Table 14 in Appendix shows that 30\% of the Sybil detection studies employ Area Under the Curve (AUC). 10 studies on sockpuppet detection techniques adopt Accuracy, Precision, F-measure and false positive rate. Accuracy is the most dominant evaluation metric in social botnet detection evaluation. Furthermore, Precision and Recall are utilized in four of the studies on the compromised account detection techniques. Identity cloning detection evaluation only relies on Accuracy, Precision and Recall.
\subsection{Datasets}
Table 15 in Appendix shows a summary of contemporary datasets for detection techniques of identity deception  in social media. As presented in the table, a considerable number of the studies apply the proposed detection approaches on datasets collected from Twitter, followed by datasets of Facebook. Twitter and Facebook are appropriate social media platforms to apply the detection techniques because of their explicit graph structures. Surprisingly, none of the literature employs Youtube datasets. 30\% of the studies that rely on Twitter datasets target social botnet. Two studies apply the proposed approaches in  datasets collected from Enron, Epinions and Renren. Uwants and HK forum, Ifeng.com and Sina Weibo are only employed by sockpuppet detection techniques.  

As shown in Table 15 in Appendix, 66\% of the studies concerning Sybil detection techniques adopt the datasets collected from Facebook, followed by 33\% of the studies adopting the datasets collected from Twitter. 
45\% of sockpuppet detection techniques are evaluated over the datasets collected from Facebook and Wikipedia. Social botnet detection techniques are only experimented over Twitter datasets. Moreover, 60\% of the compromised account detection studies are performed on Twitter datasets. 25\% of the compromised account detection techniques are carried out on the datasets collected from Foursquare and Yelp. Finally, identity cloning detection techniques are only trailed on the datasets collected from Facebook (2 studies), Linkedin and Google+ (1 study each).
\section{Open Research Challenges}
In this section, we discuss the challenges and flaws identified in the existing works of the detection techniques of social media identity deception. We categorize the open research challenges into general and specific open research challenges. 
\subsection{General Open Research Challenges}
This section presents the shared open research challenges for all the detection techniques. 
\begin{itemize}
    \item \textit{Deployment of deep learning approaches.} Traditional machine learning requires feature engineering \cite{roh2019survey}. Feature engineering is one of the most challenging steps and time-consuming. On the contrary, deep learning can automatically generate features and can save the cost of feature engineering \cite{goodfellow2016deep}. Therefore, future works are worth to investigate deep learning techniques to detect identity deception attacks in social media platforms.
    
    
    
    \item \textit{Investigation of identity-based cryptography}. None of the existing works have considered cryptography viewpoints to prevent identity deception on social media platforms except for Al‐Qurishi et al. \cite{Al-Qurishi2018a}. Al‐Qurishi et al. \cite{Al-Qurishi2018a} introduced a trapdoor using pairing-based cryptography for Sybil attack prevention. No studies to date have considered investigating identity-based cryptography to prevent other identity deception attacks. The cryptography provides secure communication of digital information. It might be in the form of a public-key system where a public key can be a user's identity such as email addresses \cite{kalyani2016survey}. Identity-based cryptography also helps social media users to find friends in a privacy-preserving way.     
    
    \item \textit{Proposing labelled datasets.} Most studies to date appear to focus on the issue of  identity deception detection on social media platforms. However, few studies have proposed labelled datasets for training and evaluating identity deception techniques. Labelled datasets would facilitate the emergence of more robust machine learning models \cite{roh2019survey}. Therefore, labelled datasets are urgently required for development of more effective models. 
    
    \item \textit{Analysis of clickstream data.} Most of the studies so far have focused  on detecting identity deception in social media based on users’ behaviours, graph-based approaches or machine learning-based approaches. Clickstream data provides information about pages' sequences or the path viewed by users when they browse a website. In identity theft, clickstreams of the attackers differ from the clickstreams made by a real owner \cite{wang2017clickstream}. Also, in sockpuppets, clickstreams of the multiple accounts might be the same. There is a strong need to propose clickstreams-based approaches to detect identity deception attacks on social media platforms.
    
    \item \textit{Development of real-time pre-attack detection approaches.} Most studies in this field so far have mostly been concerned with the detection of identity deception attacks on social media platforms after attacks occurred. However,  there are no approaches to detect identity deception attacks before any malicious behaviours taking place. The more freedom given a malicious user the higher  potential damage of the user would bring on to social media platforms \cite{Tsikerdekis2018}. For example, the pre-attack behavioural patterns of the malicious user might be discovered and analyzed based on real-time data. A user's real-time behaviour can then be monitored based on the discovered patterns. The risk of the user might be predicted before real attacks occurred. However, research has not yet considered the potential of these detection approaches to detect malicious user identities in real time. Therefore, development of real-time pre-attack detection approaches is urgently needed.
    
    
\end{itemize}

\subsection{Specific Open Research Challenges }
This section presents the open research challenges to each subcategory of detection techniques.
\subsubsection{Issues in Fake Profile Detection Techniques}
\paragraph{Issues in Sybil Detection Techniques}
\begin{itemize}
\item \textit{Designing an approach to detect Sybils before they communicate with real users.} Sybil attacks depend on the creation of multiple fake users. Therefore, there is a need to develop an approach which detects the creation of fake users before they communicate with real users. Most of the existing techniques have focused on detection of Sybil attacks instead of detecting them before they communicate with honest users. Therefore, fake users must be detected immediately following registration. Detection of fake users will increase social media security. 

\item \textit{Exploring various types of social media networks.} Most of the existing techniques focus on general social media networks such as Facebook, Twitter and LiveJournal. However, none of the existing studies applied their techniques to social shopping networks such as Polyvore, Etsy and Fancy.  Social shopping networks help users to follow different brands, to share recommendations among consumers and to make a purchase \cite{shen2012social}. Malicious accounts can reach many victims since social shopping networks have millions of users who want to buy genuine items \cite{Cao2014}. It is important to explore different kinds of social media networks. 

\item \textit{Studying the homophily strength for each edge.} Most of the current RW-based approaches need the social network to have a strong homophily property. A social platform has strong homophily property when two corresponding users are extremely probable to have the same labels \cite{Jia2017}. Existing RW-based approaches show limited detection accuracy on weak-homophily social networks.  Most researchers have focused on detecting Sybil attacks in strong homophily networks except for Sybilwalk \cite{Jia2017} and SybilSCAR \cite{Wang2018b}. Unfortunately, none of the studies to date have studied the homophily strength for each edge \cite{Wang2018b}.
\item \textit{Employment of structure, community and distribution features.} Sybil attack detection techniques mostly depend on the feature of content, user profile, time and graph. However, none of the existing techniques have employed features from new dimensions such as structure, community and distribution features. These features have been proved effective in detecting different types of identity deception attacks such as sockpuppets and social botnets.

\end{itemize}
\paragraph{Issues in Sockpuppet Detection Techniques}
\begin{itemize}
\item \textit{Exploring social media platforms with anonymous setting.} Detection techniques of sockpuppet accounts have been applied to social media platforms and online discussion forums with open settings. However,  there have been no studies made to detect sockpuppets in completely anonymous settings, for example, 4chan.  Kapoor and Gunta \cite{kapoor2016impact} discussed the behavioural effects of anonymous communication. They leveraged a real case that business reputation was ruined by an anonymous blog. Anonymous users successfully influence many social media users and stained the reputation of the business in the most serious way. Therefore, identity deception techniques should be applied to anonymous settings in social media platforms.

\item \textit{Exploring knowledge-sharing platforms.} Sockpuppet detection techniques mostly focus on social networks and discussion forums. However, sockpuppet behaviours may differ in knowledge-sharing platforms; for example, StackOverflow. Sockpuppets may aim to provide extra ``upvotes'' to their answers. Such work can produce additional insights into sockpuppet detection techniques.

\item \textit{Investigating sockpuppet behaviours.} A robust technique to identify sockpuppets may uncover an even more comprehensive variety of behaviours. For instance, sockpuppets can exist beyond single social media. Kumar et al.  \cite{Kumar2017} found 14 various groups of sockpuppet. Those groups existed in different online communities. Therefore, investigating these kinds of sockpuppets can help to characterize how the behaviours of sockpuppets may differ in various social media.

\end{itemize}
\paragraph{Issues in Social Botnet Detection Techniques}
\begin{itemize}
\item \textit{Exploring Facebook, discussion forums, Wikipedia and Instagram.} Most existing detection techniques on social botnets focus on Twitter. None of the related works consider other social media domains such as Facebook, discussion forums, Wikipedia and Instagram. Social botnets can affect different types of social media. Therefore, future research might explore other domains of social media in order to detect social botnets.

\item \textit{Usage of graph-based techniques.} Social botnet detection techniques use machine learning-based technique, text mining-based and deep learning-based techniques. However,  none of the existing techniques have attempted graph-based techniques.  Such techniques have shown efficiencies in Sybil and sockpuppet detection. Therefore, it is worthwhile to explore the application of graph-based techniques for social botnet detection.

\item \textit{Exploring social botnets in a new context such as interference of bots with public discourse.} Most of the existing works detect social botnets based on the account levels and tweet levels. Future works might examine the conversation of social media in various contexts. This would help to determine the scope of social botnets' interference with public discourse. This would also help to understand social botnets' capabilities over time.

\item \textit{Employment of user profile, time, graph and community features.} Social botnet detection techniques mostly employ content-based features. No existing detection technique has used other features.  User profile-based features, time-based features, graph-based features and community-based features have been proved effective in detecting different types of identity deception attacks.  Therefore, future research might employ and combine various features from these dimensions.

\end{itemize}

\subsubsection{Issues in Identity Theft Detection Techniques}
\paragraph{Issues in Compromised Account Detection Techniques}
\begin{itemize}
\item \textit{Employment of graph and structure-based features.} Compromised account detection techniques mostly employ content-based features, user profile-based features and time-based features. Other features such as graph-based and structure-based features have proven effectiveness in detecting other types of identity deception attacks. Therefore, future studies might focus on those new features.

\item \textit{Evaluation of friend network similarities.}  None of the existing studies have evaluated the friend network similarity except for He et al. \cite{He2014}. It relies on the number of mutual friends of two identities. However, the extent of friendships might be different in real life. For example, family members might be more important than colleagues. This indicates that people might weight the relationship value. Therefore, it is important to consider relationship weights when calculating friend network similarity. This would match the real-life situation and might make friend network similarity measure more accurate.


\end{itemize}
\subsubsection{Issues in Identity Cloning Detection Techniques}
\begin{itemize}
\item \textit{Use of machine learning-based techniques.} Studies to date have mostly been concerned the detection of identity cloning based on user profile techniques. However,  none of the existing techniques has used machine learning-based techniques. Such techniques have shown efficiency in Sybil and sockpuppet detection. Therefore, it is worthwhile to investigate the application of machine learning-based techniques.

\item \textit{Employment of content, time, graph and community features.}  Identity cloning detection techniques mostly depend on profile information. None of the existing detection techniques have explored new features.  Content-based features, time-based features, graph-based features and community-based feature have all been proved effective in detecting different types of identity deception attacks.  Therefore, future studies could employ and combine various features from different dimensions.

\item  \textit{Applying Twitter, Wikipedia, Instagram and Foursquare datasets.} Identity cloning detection techniques have been applied frequently to Facebook datasets. Identity cloning detection techniques need to be applied to other social media platforms such as Twitter, Wikipedia, Instagram and Foursquare in order to evaluate the generalization of those techniques. 
\end{itemize}
\section{CONCLUSION}
Social media identity deception detection is a fast-growing research area and has received a considerable amount of attention from both researchers and the IT industry.  Meanwhile, many scholarly contributions have been made to identity deception detection. In this paper, we conducted a comprehensive analysis of the literature on that subject. First, we categorized identity deception into three major attacks: 1) fake profile, which includes Sybil attacks, sockpuppets and social botnets; 2) identity theft, which includes compromised and hijacked accounts and 3) identity cloning which includes single-site cloning and cross-site cloning. We then made a comprehensive analysis of each identity deception attack. We also provided a comprehensive review of existing detection techniques on identity deception in social media. Finally, we explored future research directions for each type of identity deception detection techniques.
\section*{SUPPLEMENTARY MATERIALS}
All comparison tables are available in the online version of this paper.

\bibliographystyle{ACM-Reference-Format}
\bibliography{sample-base}

\appendix

\end{document}